\documentclass[a4paper,12pt]{article}

\usepackage{jheppub}

\usepackage[english]{babel}
\pdfoutput=1
\usepackage[utf8]{inputenc}
\usepackage[T1]{fontenc}
\usepackage{amsmath}
\usepackage{amssymb}
\usepackage{float}
\usepackage{graphicx}
\usepackage{comment}
\usepackage[dvipsnames]{xcolor}
\usepackage{color}
\usepackage{epsfig}
\usepackage{dcolumn}

\usepackage{bm}
\usepackage[caption=false]{subfig}
\usepackage{epstopdf}
\usepackage{setspace}
\usepackage[normalem]{ulem}	
\usepackage{multicol}
\usepackage{mathrsfs}
\usepackage{cancel}
\usepackage{fancybox}
\pdfoutput=1
\usepackage{braket}

\hypersetup{
	unicode = true,
	pdfborder = {0 0 0},
	linktoc = page,
	colorlinks,
	linkcolor = blue,
	citecolor = black,
	filecolor = black,
	urlcolor = blue
}

\newcommand{\vEW}{v_\text{\tiny EW}}

\definecolor{colRed0}{rgb}{0.85, 0.05, 0.12}
\definecolor{colRed1}{rgb}{0.92, 0.1, 0.05}
\definecolor{colRed2}{rgb}{0.95, 0.35, 0.05}
\definecolor{colYellow1}{rgb}{1., 0.82, 0.}
\definecolor{colBlue1}{rgb}{0.0, 0., 0.4}
\definecolor{colBlue2}{rgb}{0.1, 0.3, 0.9}
\definecolor{colBlue3}{rgb}{0.15, 0.4, 0.75}
\definecolor{colBlue4}{rgb}{0.3, 0.8, 0.93}
\definecolor{colGreen0}{rgb}{0.0, 0.15, 0.05}
\definecolor{colGreen1}{rgb}{0.0, 0.35, 0.1}
\definecolor{colGreen2}{rgb}{0.1, 0.65, 0.2}
\definecolor{colGreen3}{rgb}{0.3, 0.85, 0.5}
\definecolor{colBrown1}{rgb}{0.3, 0.18, 0.12}
\definecolor{colBrown2}{rgb}{0.5, 0.3, 0.20}
\definecolor{colViolet1}{rgb}{0.4, 0.18, 0.42}
\definecolor{colViolet2}{rgb}{0.5, 0.3, 0.70}

\newcommand{\ew}{\textnormal{\tiny EW}}

\newcommand{\GeV}{\text{ GeV}}

\newcommand{\tamp}{t_\mathrm{amp}}
\newcommand{\tfrag}{\Delta t_\mathrm{frag}}
\newcommand{\phifrag}{\Delta\phi_\mathrm{frag}}
\newcommand{\phiscan}{\Delta\phi_\mathrm{scan}}
\newcommand{\tnl}{t_\mathrm{nl}}

\newcommand{\deltakcr}{\delta k_\mathrm{cr}}
\newcommand{\kcr}{k_\mathrm{cr}}

\title{\textbf{Axion Fragmentation on the Lattice}}

\date{\today}

\author[a]{Enrico Morgante,}
\author[a]{Wolfram Ratzinger,}
\author[b]{Ryosuke Sato,}
\author[c]{Ben A. Stefanek}

\affiliation[a\,]{PRISMA$^+$  Cluster  of  Excellence  and  Mainz  Institute  for  Theoretical  Physics, Johannes  Gutenberg-Universit\"at  Mainz,  D-55099  Mainz,  Germany}
\affiliation[b\,]{Tsung-Dao Lee Institute, Shanghai Jiao Tong University, 800 Dongchuan Road, Shanghai 200240, China}
\affiliation[c\,]{Physik-Institut, Universit\"at Z\"urich, CH-8057 Z\"urich, Switzerland}

\emailAdd{emorgant@uni-mainz.de}
\emailAdd{w.ratzinger@uni-mainz.de}
\emailAdd{ryosukesato64@gmail.com}
\emailAdd{bestef@physik.uzh.ch}

\abstract{
We analyze the phenomenon of axion fragmentation when an axion field rolls over many oscillations of a periodic potential. This is particularly relevant for the case of relaxion, in which fragmentation provides the necessary energy dissipation to stop the field evolution. We compare the results of a linear analysis with the ones obtained from a classical lattice simulation, finding an agreement in the stopping time of the zero mode between the two within an ${\cal O}(1)$ difference. We finally speculate on the generation of bubbles with different VEVs of the axion field, and discuss their cosmological consequences.
}

\begin{document}

\begin{flushright}
MITP-21-045
\end{flushright}

\maketitle

\section{Introduction}
Axion-like particles (ALPs) are a common feature of beyond the Standard Model (BSM) physics, arising as pseudo Nambu-Goldstone bosons of spontaneously broken global symmetries. While the most renowned application is the axion solution to the strong CP problem~\cite{Peccei:1977hh, Peccei:1977ur, Weinberg:1977ma, Wilczek:1977pj}, ALPs routinely find their place in BSM physics as, \textit{e.g.}, natural dark matter (DM)~\cite{Abbott:1982af, Dine:1982ah,Preskill:1982cy,Hui:2016ltb} and inflaton candidates~\cite{Freese:1990rb,Pajer:2013fsa, Adshead:2015pva, Domcke:2019qmm, Adshead:2019lbr}, or as generic particles in the low-energy spectrum of string theories with a wide range of possible masses~\cite{Witten:1984dg,Svrcek:2006yi,Arvanitaki:2009fg,Marsh:2015xka}. Relatively typical string constructions can also result in an explicit breaking of the discrete shift symmetry of ALP, leading to an axion monodromy~\cite{McAllister:2008hb,Silverstein:2008sg,Marchesano:2014mla,Blumenhagen:2014gta,Hebecker:2014eua,McAllister:2014mpa}. In the presence of a monodromy, the ALP field space is no longer compact and the ALP can probe multiple non-degenerate minima of the potential. This effect was exploited in the context of inflation to allow for potentials which slowly varied over a super-Planckian field range~\cite{McAllister:2008hb,Silverstein:2008sg,Marchesano:2014mla}. 

It was realized such a potential can also provide a novel dynamical solution to the electroweak (EW) hierarchy problem~\cite{Graham:2015cka}. The mechanism relies on the cosmological evolution of an axion-like field $\phi$, called the \emph{relaxion}, with a softly-broken discrete shift symmetry allowing a monodromy-like term in the potential as well as a linear coupling to the SM Higgs field. As the relaxion rolls down its potential, the Higgs mass decreases from a natural value of order of the cutoff scale of the theory until it becomes close to zero and a stopping mechanism of one type or another is triggered. The original model occurs during inflation and identifies the relaxion with the QCD axion, where non-perturbative QCD instanton effects generate a periodic potential for $\phi$ in the presence of non-vanishing quark masses. The appearance of the barriers as the Higgs mass scans through the origin and triggers a non-zero EW vacuum expectation value (VEV) acts as the trigger that stops the relaxion field, with Hubble friction and a small, technically natural slope ensuring the correct EW VEV is not overshot. Issues with the original model were addressed in~\cite{Espinosa:2015eda} and alternative stopping mechanisms using, \textit{e.g.}, friction from the tachyonic production of EW gauge bosons~\cite{Hook:2016mqo, Fonseca:2018xzp, Fonseca:2018kqf, Fonseca:2020pjs},%
\footnote{In this case,  a population of relaxion particles is produced after reheating via  scattering,  with an abundance that matches the observed DM one for a relaxion mass in the keV range~\cite{Fonseca:2018kqf, Fonseca:2020pjs}.
Such a scenario is anyway strongly constrained by structure formation probes~\cite{Baumholzer:2020hvx}.}
friction from parametric resonance due to the Higgs zero mode~\cite{Ibe:2019udh}, potential instabilities~\cite{Wang:2018ddr}, fermion production~\cite{Kadota:2019wyz}, and dark photon production \cite{Choi:2016kke, Domcke:2021yuz} have also been considered.

Not much attention has been focused on the role of ALP quantum fluctuations, which can be excited by the cosmological evolution of the homogeneous zero mode. In typical situations where the ALP oscillates about a single minimum, fluctuation growth is suppressed unless the amplitude is large enough for anharmonic corrections to the potential to be important~\cite{Greene:1998pb, Arvanitaki:2019rax}. However, even harmonically oscillating ALPs can nonetheless excite other coupled degrees of freedom, such as other scalar fields~\cite{Dolgov:1989us, Traschen:1990sw, Kofman:1994rk, Shtanov:1994ce, Kofman:1997yn} or gauge field quanta through a Chern-Simons type coupling~\cite{Cook:2011hg,Barnaby:2011qe,Barnaby:2011vw,Pajer:2013fsa,Domcke:2019qmm,Machado:2018nqk,Machado:2019xuc,Ratzinger:2020oct}, possibly also resulting in the production of gravitational waves (GWs)~\cite{Banerjee:2021oeu}.

The situation for an ALP with a non-compact potential and sufficient velocity to overcome many barriers is different. As the relaxion or ALP rolls down its potential, it has a highly oscillatory mass term as it traverses a large number of fundamental periods. This generically results in a parametric resonance effect that leads to exponential growth of fluctuations in the ALP field for a particular range of momenta~\cite{Fonseca:2019ypl}. This so-called fragmentation of the ALP field results in friction that can stop the field as kinetic energy is transferred from the homogeneous zero mode into ALP quanta. This effect leads to a natural and novel stopping mechanism for the relaxion, the so-called \emph{self-stopping relaxion}, as first pointed out in~\cite{Fonseca:2019ypl,Fonseca:2019lmc}. Similar self-resonance effects have been considered in the context of
%oscillons~\cite{ Hertzberg:2010yz, Amin:2010xe, Amin:2010dc, Antusch:2017flz, Olle:2019kbo,Sang:2019ndv,Berges:2019dgr}
axion monodromy inflation \cite{Flauger:2009ab} and axion monodromy dark matter \cite{Jaeckel:2016qjp, Berges:2019dgr} and can also result in GW production~\cite{Chatrchyan:2020pzh}.

The necessary ingredients for successful relaxation of the EW scale in the context of the self-stopping relaxion were studied in~\cite{Fonseca:2019lmc}, while Ref.~\cite{Fonseca:2019ypl} examined the conditions under which ALP fragmentation can efficiently stop the field evolution for generic ALPs. In particular, the time required to stop the field as well as the corresponding field displacement were computed in a linearized analysis, where the equation of motion for the ALP fluctuations can be Fourier transformed into momentum space, with each mode evolving independently.
It was shown in~\cite{Fonseca:2019ypl} that the linear approximation holds for most of the fragmentation process, thus the linearized results for \textit{e.g.}, the stopping time were expected to hold up to $\mathcal{O}(1)$ corrections from non-linearities. 

A fully satisfactory description of the system in the non-linear regime requires a detailed lattice study, which we perform in this work. While our motivation is rooted in the relaxion mechanism, our lattice study here is broadly applicable to general ALPs with/without a monodromy-like potential. In particular, another interesting example for the application of the axion fragmentation would be the kinetic misalignment scenario \cite{Co:2019jts, Chang:2019tvx, DiLuzio:2021gos}, which is a novel ALP dark matter production mechanism. In this scenario, ALP zero mode has initial velocity which is large enough to overcome the potential barrier, and this initial velocity determines the amount of the relic abundance today. Physical consequences of the axion fragmentation in the ALP dark matter scenario will be discussed elsewhere \cite{ALPDMfragmentation}. We solve the ALP equations of motion in position space on a discretized spacetime lattice using a staggered grid algorithm~\cite{Figueroa:2017qmv,Cuissa:2018oiw}, which reproduces the continuum version of the equations up to an error that is quadratic in the lattice spacing. We generically find that ALP fragmentation is more efficient in the presence of non-linearities, mainly due to the importance of $2\rightarrow 1$ processes that allow for the growth of modes outside the parametric resonance band. The more efficient fragmentation typically leads to an order of magnitude reduction in the stopping time and field displacement as compared to the results from the linear analysis. As expected, the final ALP spectrum is broadened compared to the linearized analysis, and the final field configuration is highly inhomogeneous as most of the energy in the system is contained in fluctuations corresponding to axion particles. The rest of the features of the linear analysis are qualitatively confirmed, and we comment briefly on the possible formation of domain walls due to the ALP field stopping in different minima on scales separated by more than the inverse stopping time.

%Introduction Outline
%---------------------------
%\begin{itemize}
%\item ALP motivation
%\item Monodromy
%\item Relaxion solution to the hierarchy problem
%\item Self-stopping relaxion
%\item Need for lattice sim, motivation is relaxion but focus here on general ALPs
%\end{itemize}

\section{Summary of the linear analysis}
\label{sec:linearsummary}

In this section we briefly recall the results of Ref.~\cite{Fonseca:2019ypl}.
We consider a potential of the form
\begin{equation}\label{eq:axion potential}
V (\phi)= - \mu^3\phi + \Lambda_b^4  \cos\frac{\phi}{f} \,,
\end{equation}
where we assume $\Lambda_b^4/f \gg \mu^3$ and we define the axion mass $m^2 = \Lambda_b^4/f^2$.  Notice that, for $\Lambda_b^4/f \sim \mu^3$, the physical value of the axion mass is smaller than this value. This does not affect our discussion. In the rest of the paper we will use $m$ or $\Lambda_b$ interchangeably.
We assume that the axion has an initial kinetic energy large enough to overcome the barriers of the potential, $\dot\phi^2/2 \gg \Lambda_b^4$.

We decompose the axion field into a homogeneous mode plus small fluctuations:
\begin{equation}
\phi(x,t) = \phi(t) + \delta\phi(x,t) = \phi(t) + \left( \int \frac{d^3k}{(2\pi)^3}a_k u_k(t) e^{ikx} + h.c. \right)\label{eq:ansatz}
\end{equation}
where $a_k$ are the usual annihilation operators with $[a_k,a_{k'}^\dagger] = (2\pi)^3\delta^3(k-k')$.
As initial condition, we assume the modes are initially in the Bunch-Davies vacuum,
where $u_k(t) \approx e^{-i k \tau}/(a\sqrt{2k})$ with $\tau$ being the conformal time.\footnote{
Note that Eq.~(2.6) in Ref.~\cite{Fonseca:2019ypl} contains an error in the phase of the Bunch-Davies mode functions, which does not affect the derivation of the subsequent results. In addition, the effects of cosmic expansion are not important since fragmentation is much faster than one Hubble time.
}
The equations of motion for the zero mode $\phi(t)$ and for the mode functions $u_k$ are given by
\begin{align}
\ddot{\phi} + 3H\dot{\phi} + V'(\phi) + \frac{1}{2}V'''(\phi) \int\frac{d^3 k}{(2\pi)^3} |u_k|^2 &=0, \label{eq:zeromode}\\
\ddot{u_k} + 3H\dot{u_k} + \left[ \frac{k^2}{a^2} + V''(\phi)\right] u_k &= 0. \label{eq:fluctuation}
\end{align}
Equation~(\ref{eq:zeromode}) is such that a growth of the mode functions $u_k$ slows down the evolution of the zero mode $\phi$.
Neglecting cosmic expansion, and in the limit of constant velocity, Eq.~(\ref{eq:fluctuation}) can be read as a Mathieu equation, which features exponentially growing solutions depending on its parameters, namely when $k$ falls in specific bands around $n \dot\phi /(2f)$, for integer $n\geq 1$. Modes falling in the $n=1$ modes grow faster, and the width of the band is larger than for $n\geq 2$, thus we expect these modes to be the principal source of friction to the axion.

For $\dot\phi^2/2\gg\Lambda_b^4$, the $n=1$ instability band can be written as $|k-k_\mathrm{cr}|<\deltakcr$, with
\begin{equation} \label{eq:kcr and delta kcr}
k_\mathrm{cr} = \frac{\dot\phi}{2f} \, ,
\qquad
\deltakcr = \frac{\Lambda_b^4}{2f \dot\phi}\,.
\end{equation}
The asymptotic behaviour of $u_k$ at large $t$ is
\begin{equation}
%u_k \propto \exp\left( \sqrt{ (\delta k_{\rm cr} )^2 - \left( k- k_{\rm cr}  \right)^2 } t \right) \sin \left(k_{\rm cr} t - \arctan\sqrt{ \frac{\delta k_{\rm cr} - (k - k_{\rm cr} ) }{ \delta k_{\rm cr} + (k - k_{\rm cr} )  } } \right). \label{eq:asymptotic uk}
u_k \sim (2k_{\rm cr})^{-1/2} \exp\left( \sqrt{ (\delta k_{\rm cr} )^2 - \left( k- k_{\rm cr}  \right)^2 } t \right) \sin\left( k_{\rm cr} t + \frac{\pi}{4}\right) . \label{eq:asymptotic uk}
\end{equation}
Due to this exponential growth, the energy density of the fluctuations within the instability band increases. Energy conservation implies that the kinetic energy of the zero mode decreases by the same amount, thus reducing $\dot\phi$ and correspondingly shifting the instability band towards smaller $k$'s. At the linear level, the growth of the modes around $k_\mathrm{cr}$ stops when they exit the instability band, \textit{i.e.}, when the critical mode has decreased by an amount $\deltakcr$. As we will discuss in the following in Sec.~\ref{sec:NLO}, at next to leading order the scattering of two modes of the instability band can enhance modes which are still outside the latter. As a result, these modes enter into the instability band with a larger initial amplitude. Hence, the time needed for their enhancement to level which induces a significant backreaction is shortened, increasing the overall efficiency of the process.

The equation of motion of the fluctuations Eq.~(\ref{eq:fluctuation}) can be solved, assuming $\ddot\phi$ does not vary during the amplification time of a single mode, by means of a WKB approximation in three separate time intervals: first, before the mode $k_\mathrm{cr}$ enters the instability band; second, when the mode is deep inside the instability band; third, after it has left it. In the two transition regions, when the mode enters and exits the instability band, the solution can be expressed in terms of Airy functions. Continuity of the solution is then used to match the five intervals.
The asymptotic solution for $u_k$, after it has left the instability band, is found to be
\begin{align}
u_{k_\mathrm{cr}}(t) \simeq \frac{1}{a} \sqrt{\frac{2}{k_\mathrm{cr}}} \exp\left( \frac{\pi \Lambda_b^8}{4f\dot\phi^2\,|\ddot\phi + H\dot\phi|} \right) \sin\left( \frac{1}{a} k_\mathrm{cr} t + \delta \right)\,,
\label{eq:asymptotic uk 2}
\end{align}
and the time needed for this amplification is
\begin{equation}\label{eq:delta t amp}
\delta \tamp \approx \frac{1}{2\deltakcr} \log \frac{\dot\phi^2}{k_\mathrm{cr}^4}
= \frac{f \dot\phi}{\Lambda_b^4} \log \frac{16 f^4}{\dot\phi^2} \,.
\end{equation}
By using energy conservation and Eq.~(\ref{eq:asymptotic uk 2}), the equation of motion for the zero mode can be derived:
\begin{align}
\dot\phi\ddot\phi
=
- 3 H \dot\phi^2
+ \mu^3 \dot\phi
- \frac{1}{32\pi^2 f^4} \dot\phi^3 |\ddot\phi + H\dot\phi| \exp\left( \frac{\pi \Lambda_b^8}{2f\dot\phi^2|\ddot\phi + H\dot\phi|} \right) \,.
\label{eq:eq for phi double dot}
\end{align}
Equation~(\ref{eq:eq for phi double dot}) can be integrated exactly for $H = 0$, $\mu = 0$. In particular, one finds that the evolution of the zero mode is stopped by the backreaction after a time
\begin{equation}
\label{eq:mastertimescale}
\Delta t_{\rm frag} \simeq \frac{2 f \dot{\phi}_0^3}{3 \pi \Lambda_b^8}\log\frac{32 \pi^2 f^4}{\dot{\phi}_0^2} \,,
\end{equation}
and the corresponding field excursion is
\begin{equation}
\label{eq:masterfieldexcursion}
\Delta \phi_{\rm frag} \simeq \frac{ f \dot{\phi}_0^4}{2 \pi \Lambda_b^8}\log\frac{32 \pi^2 f^4}{\dot{\phi}_0^2} \,.
\end{equation}
The effect of Hubble friction and of the slope $\mu$ is negligible as long as the following equation is satisfied:
\begin{equation}
\mu^3 < 
2H\dot\phi_0 + \displaystyle\frac{\pi\Lambda_b^8}{2f\dot\phi_0^2} \left( W_0\left( \displaystyle\frac{32\pi^2f^4}{e\dot\phi_0^2} \right) \right)^{-1}\,.
\label{eq:condition for no positive phiddot}
\end{equation}
Here $W_0(z)$ is the 0-th branch of the product logarithm function.
If the slope $\mu$ is too large, the field is accelerated and the fragmentation is not efficient enough to stop it, unless Hubble friction balances it. In Sec.~\ref{sec:lattice} we will check the validity of Eqs.~(\ref{eq:mastertimescale})--(\ref{eq:condition for no positive phiddot}) with a lattice analysis. Due to the increased efficiency at next-to-leading order (NLO), the time scale and the field excursion of Eqs.~(\ref{eq:mastertimescale}),~(\ref{eq:masterfieldexcursion}) are reduced typically by a factor of a few. Instead, Eq.~(\ref{eq:condition for no positive phiddot}) is satisfied with order percent accuracy.

\section{Lattice analysis}\label{sec:lattice}

The linear analysis presented above is very useful as it provides simple analytic expressions for the quantities related to the axion evolution. One may wonder, though, whether these results are robust once non-linear effects are taken into account. Even though a strong backreaction is intrinsically related to a breakdown of perturbativity, it is expected that, at NLO, the efficiency of fragmentation is not suppressed in a potential as in Eq.~(\ref{eq:axion potential})~\cite{Fonseca:2019ypl}. In this section, we discuss the validity of this statement by means of a lattice simulation.

The simulation is carried out using a staggered grid quantization of space and time, guaranteeing second order accuracy in the lattice spacing $\mathcal{O}(dx_{\mu}^2)$. The time integration of the resulting field equations is carried out using a leapfrog algorithm (see \cite{Figueroa:2020rrl} for a recent review of lattice techniques). We vary the side length of the simulated box $L$ as well as the number of lattice sites $N$ to ensure that our results are independent of them, which is the case as long as the critical modes from the start of the simulation when $\langle\dot\phi\rangle=\dot\phi_0$ up to the end where $\langle\dot\phi\rangle< 2 m f$ are all covered. This corresponds to $dx=L/\sqrt[3]{N}\ll 2f/\dot\phi_0$ and $L\gg 1/m$.

We start neglecting the slope and cosmic expansion, such that $\mu=H=0$. The most relevant quantities that we want to compute on the lattice are the duration of and the field excursion during the fragmentation process.
From the linear analysis, we know that the modes that are inside of the first instability band at the time when the barriers appear, will grow for a time $\delta \tamp|_{\dot\phi=\dot\phi_0}$ as in Eq.~(\ref{eq:delta t amp}), where $\dot\phi_0$ is the initial velocity. After that time, the instability band moves towards lower $k$ modes due to the backreaction onto the zero mode. 
We are interested in the time needed to stop the evolution of the zero mode and the corresponding field excursion, which were computed in the linear approximation in  Eqs.~(\ref{eq:mastertimescale}) and~(\ref{eq:masterfieldexcursion}) to be
\begin{equation}
\Delta t_{\rm frag} \simeq \frac{2 f \dot{\phi}_0^3}{3 \pi \Lambda_b^8}\log\frac{32 \pi^2 f^4}{\dot{\phi}_0^2} \,,
\qquad
\Delta \phi_{\rm frag} \simeq \frac{ f \dot{\phi}_0^4}{2 \pi \Lambda_b^8}\log\frac{32 \pi^2 f^4}{\dot{\phi}_0^2} \,,
\label{eq:linearStopping}
\end{equation}
where, for the typical relaxion parameters, we find $2/(3\pi) \log(\dots) \sim \mathcal{O}(10)$. Let us also define the quantities
\begin{equation}
t_\mathrm{nl}  = \frac{f \dot\phi^3_0}{\Lambda_b^8} \,, \qquad
\phi_\mathrm{nl}  = \frac{f \dot\phi^4_0}{\Lambda_b^8} \,,
\end{equation}
which control the time and the corresponding distance in field space it takes for the field to come to a complete stop after fluctuations become non-linear.
At the non-linear level, we generalize the relations in Eq.~(\ref{eq:linearStopping}) via the following parameterization
\begin{equation}
\Delta t_\mathrm{frag}^\mathrm{nl} = \delta \tamp + t_\mathrm{nl} \cdot z_t \,,
\label{eq:delta t frag nl}
\end{equation}
and
\begin{equation}
\Delta \phi_\mathrm{frag}^\mathrm{nl} = \dot\phi_0 \delta \tamp + \phi_\mathrm{nl} \cdot z_\phi \,.
\label{eq:delta phi frag nl}
\end{equation}
\begin{figure}
\centering
\includegraphics[width=\textwidth]{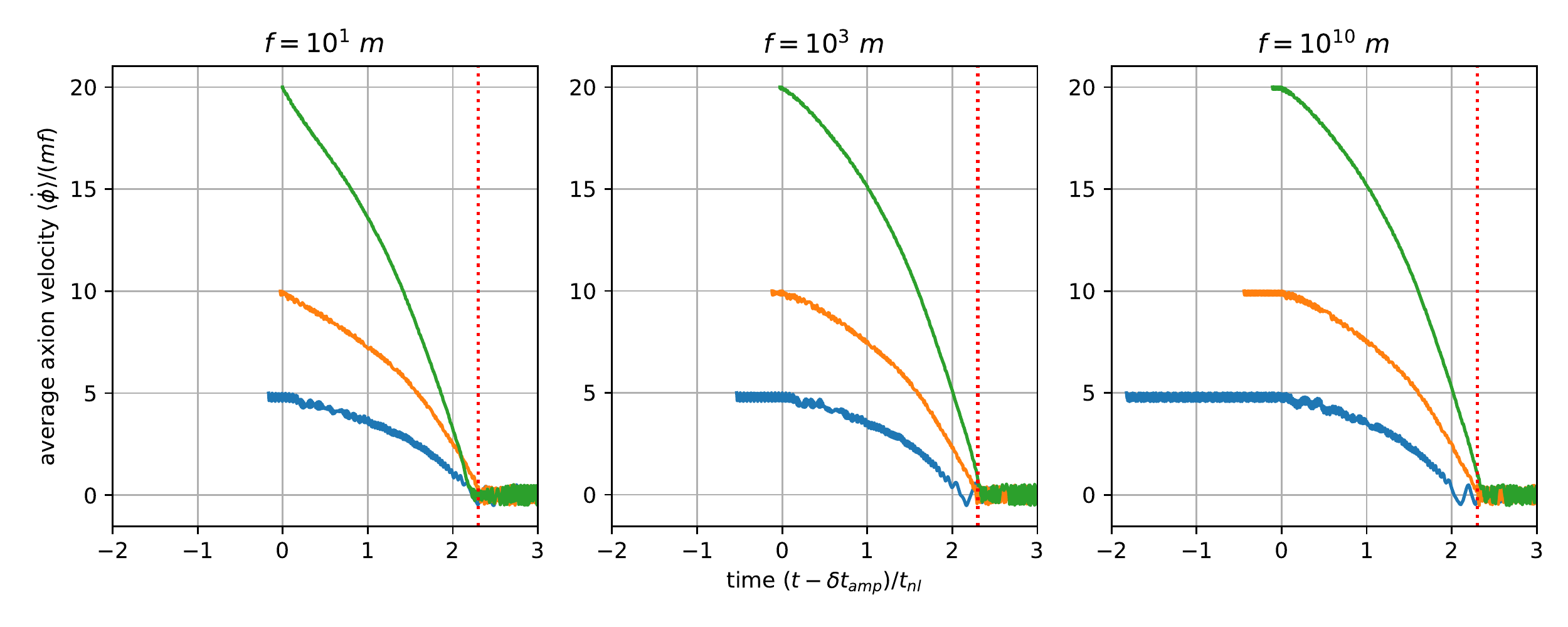}
\includegraphics[width=\textwidth]{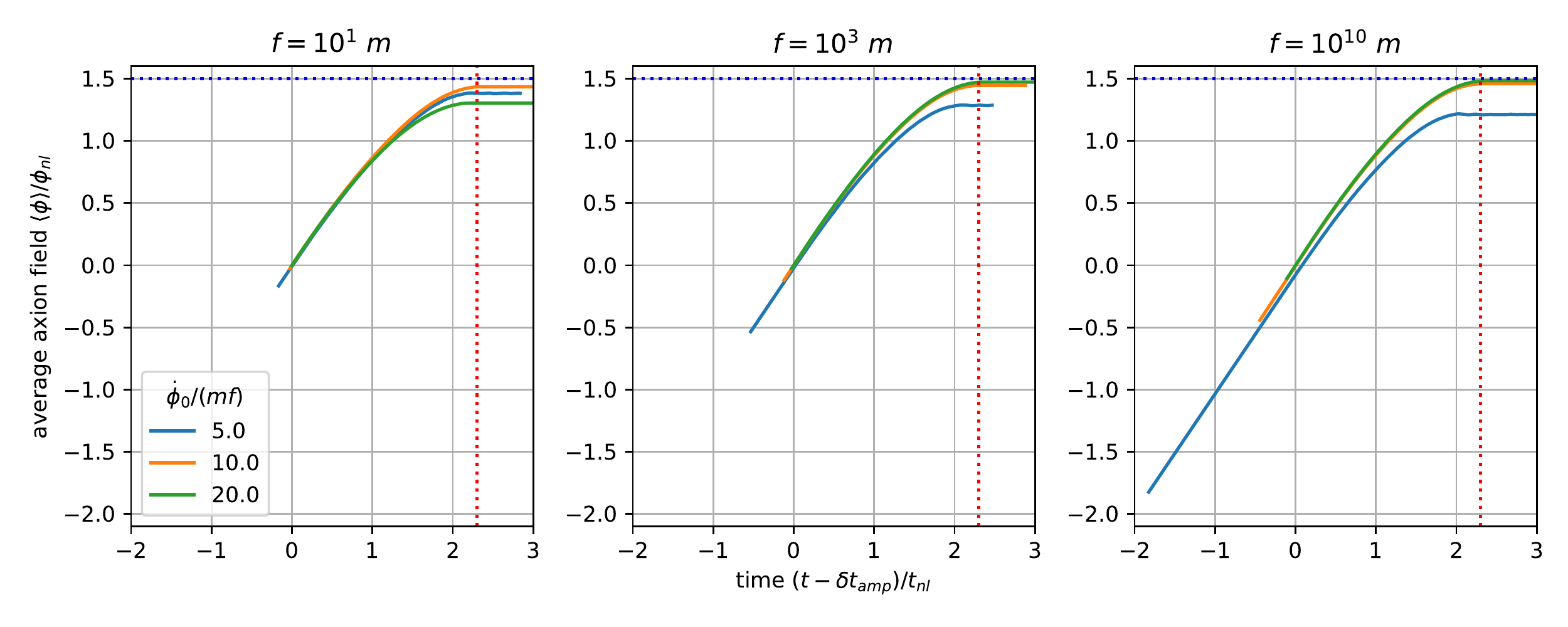}
\caption{\label{fig:AxFragScanMu0H0}
Field evolution with slope $\mu=0$ and no expansion for different initial velocities and decay constants $f$.  All simulations were run with $N=128^3$ lattice sides and length $L=20/m$ along each side. Top: We clearly see how the stopping process consists of two parts  i) a phase where the modes that are initially enhanced by parametric resonance grow from vacuum to an energy density $\rho\approx m^2f^2$ in a time $\delta \tamp$ and ii) a nonlinear part that lasts a time of $2.3\, t_{nl}$ (marked by the red dotted line).  Bottom: We see that in the non-linear regime the fields roll a distance $\approx\phi_{nl}\cdot1.5$ (blue dashed line), in the limit of large $f$ and $\dot\phi_0$.}
\end{figure}
We show in Fig.~\ref{fig:AxFragScanMu0H0} the evolution of $\dot\phi(t)$ (top) and $\phi(t)$ (bottom) for different choices of the initial velocity and of the potential parameters. It can be seen that after the short time $\delta \tamp$ in which the axion evolves with an almost constant velocity, the field slows down and stops in a time given in Eq.~(\ref{eq:delta t frag nl}) with
\begin{equation}
z_t \approx 2.3 \,,
\end{equation}
for $10m \leq f \leq 10^{10}m$ and $5mf \leq \dot\phi_0 \leq 20mf$. Analogously, in the bottom panel we see that 
\begin{equation}
z_\phi \approx 1.5 \, .
\end{equation}
These values are shorter by a factor of $\mathcal{O}(10)$ than the ones obtained in the linear analysis. 
The reason for this enhanced efficiency found in the lattice analysis is mainly due to the NLO correction that will be discussed in detail in Sec.~\ref{sec:NLO}.
This difference has a minor impact on the analysis of the relaxion parameter space of Ref.~\cite{Fonseca:2019lmc} (in which an order of magnitude uncertainty is always assumed), as we will discuss more in Sec.~\ref{sec:relaxion params}.

\begin{figure}
\centering
\includegraphics[width=0.8\textwidth]{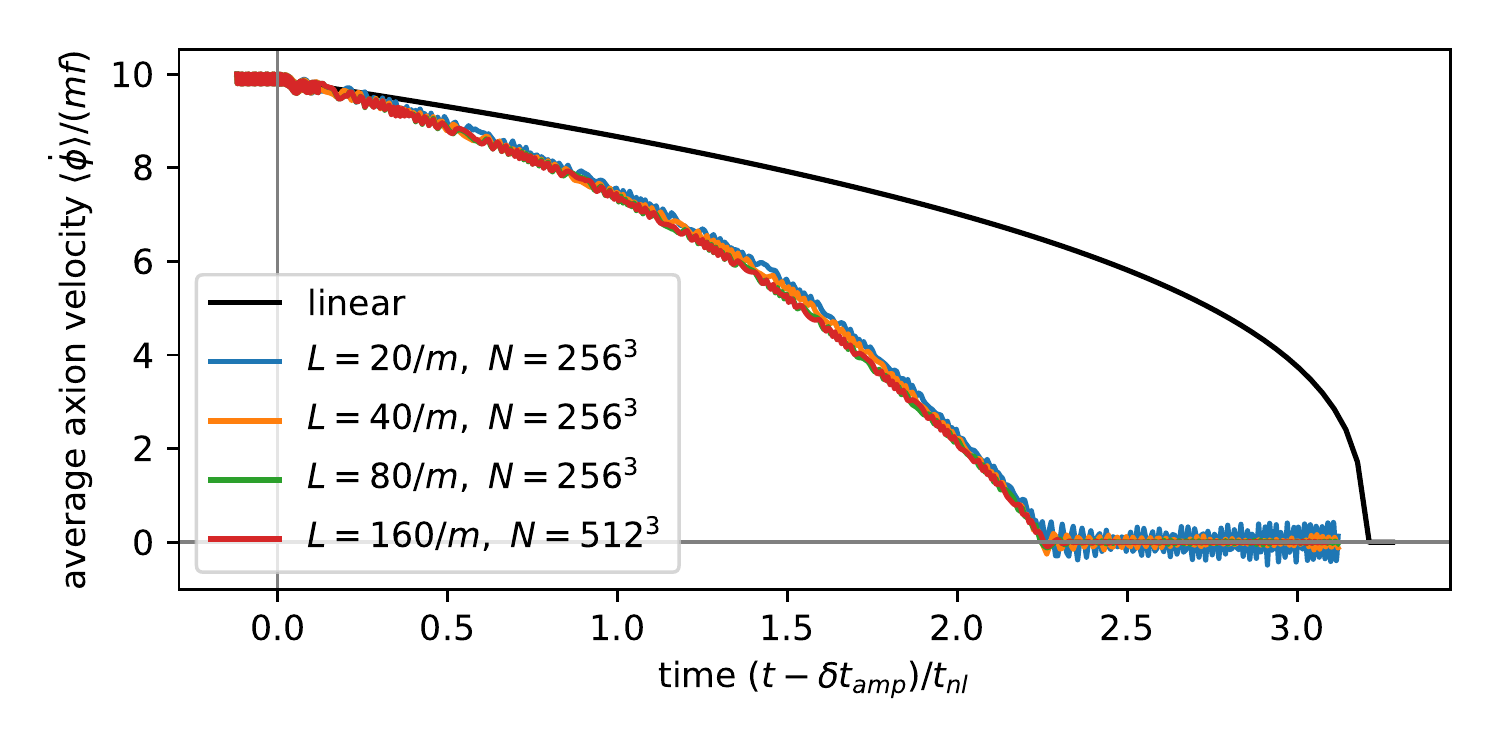}
\caption{\label{fig:lattice params}
Average axion velocity for $f=10^3\ m ,\ \mu=0$, and no expansion as obtained from linear analysis (Eq.~(\ref{eq:eq for phi double dot})) and from different realizations of the lattice. }
\end{figure}

In Fig.~\ref{fig:lattice params} we show the evolution of the axion field for different choices of the lattice parameters, which demonstrates the stability of our results.

\begin{figure}
\centering
\includegraphics[width=.7\textwidth]{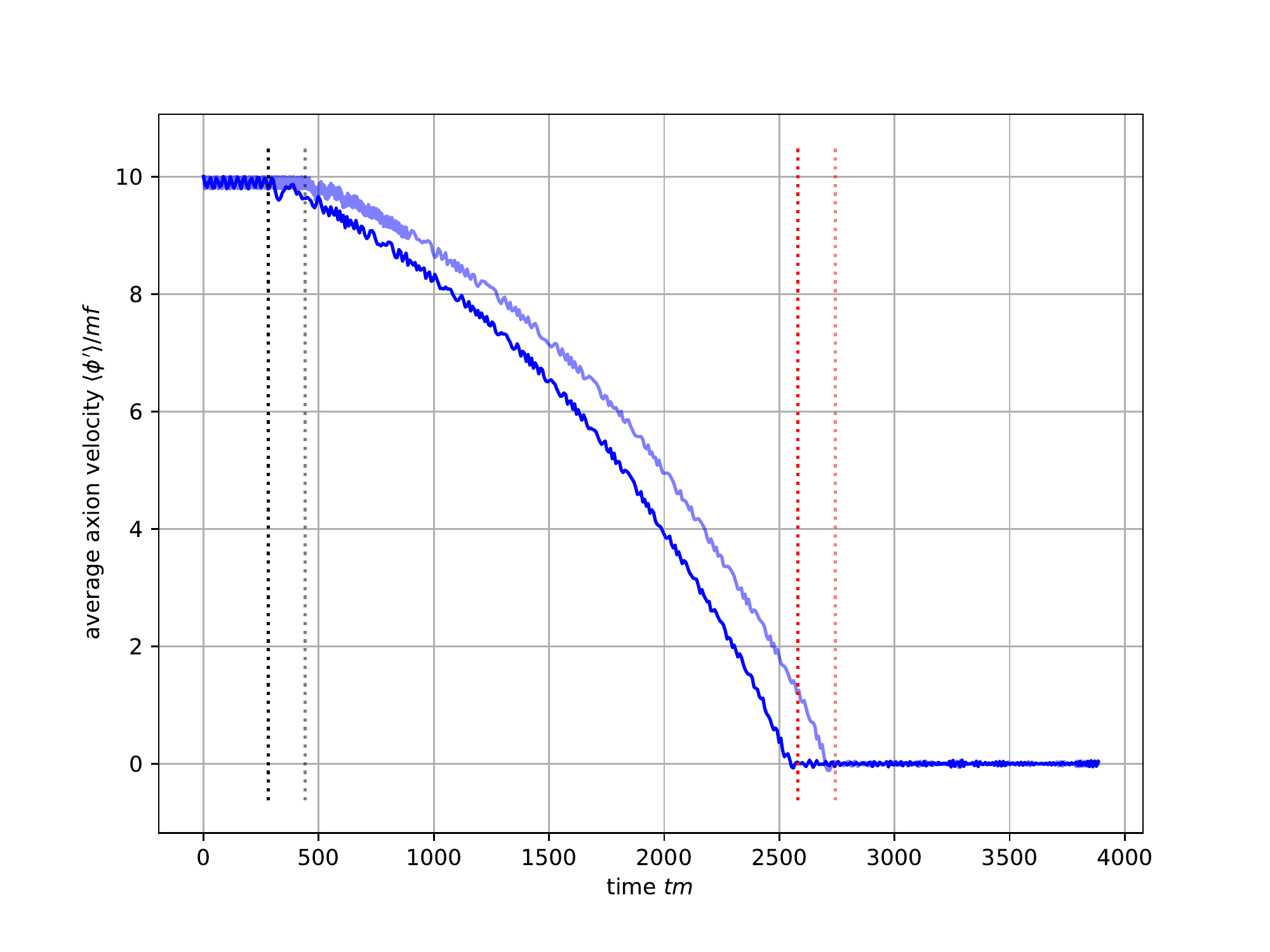}
\caption{\label{fig:flat vs Bunch Davies}
Evolution of the axion field with $\mu=0$, $f=10^{10}m$, $\dot\phi_0=10\, mf$, and no expansion for different initial energy spectra. The dark colors correspond to a flat initial energy spectrum (as expected if fluctuations are enhanced during inflation) where the energy in the initial resonance band is enhanced by a factor $d\rho/d\log k(k_{cr,0})\approx10^8\,  d\rho_{BD}/d\log k(k_{cr,0})$ as compared to the Bunch-Davies vacuum (light colors).
The gray and black dashed lines mark $\delta \tamp$ and $\delta \tamp^\mathrm{mod}$, respectively (see Eq.~(\ref{eq:delta t amp mod})), while the thin and thick red dashed lines correspond to $\delta \tamp + z_t t_\mathrm{nl}$ and $\delta \tamp^\mathrm{mod} + z_t t_\mathrm{nl}$. Both simulations were run with $N=256^3$ lattice sides and length $L=80/m$ along each side.}
\end{figure}

\medskip
If the fragmentation process takes place after inflation, one may expect the fluctuations to be enhanced during inflation compared to the Bunch-Davies spectrum and be frozen until they re-enter the horizon, with a nearly scale-invariant power spectrum. 
In Fig.~\ref{fig:flat vs Bunch Davies} we show the axion evolution in a run with an initially flat power spectrum, compared to one with the Bunch-Davies spectrum. We fix the normalization of the flat power spectrum in such a way that in the initial resonance band the power spectrum is enhanced with respect to the Bunch-Davies case by $(d\rho/d\log k)_{k_{cr,0}}\approx x  \times (d\rho_{BD}/d\log k)_{k_{cr,0}}$, and we take $x= 10^8$ in Fig.~\ref{fig:flat vs Bunch Davies}.  As it can be seen from the figure, the only difference in this case is in the duration of the amplification time $\delta \tamp$, which now lasts
\begin{align}\label{eq:delta t amp mod}
\delta \tamp  \rightarrow \,\,  \delta \tamp^\mathrm{mod} \equiv\, \frac{f \dot\phi}{\Lambda_b^4} \log \left( x \times \frac{16 f^4}{\dot\phi^2} \, \right) \,,
\end{align}
as one would expect. The duration of the non-linear phase $t_\mathrm{nl} z_t$ is instead independent of the initial power spectrum. We expect this behavior to not depend on the choice of the power spectrum, but only on the normalization of the initial instability band. This is due to the dominance of induced secondary fluctuations, as will be discussed below in Sec.~\ref{sec:NLO}.

\bigskip

\begin{figure}
\centering
\includegraphics[width=\textwidth]{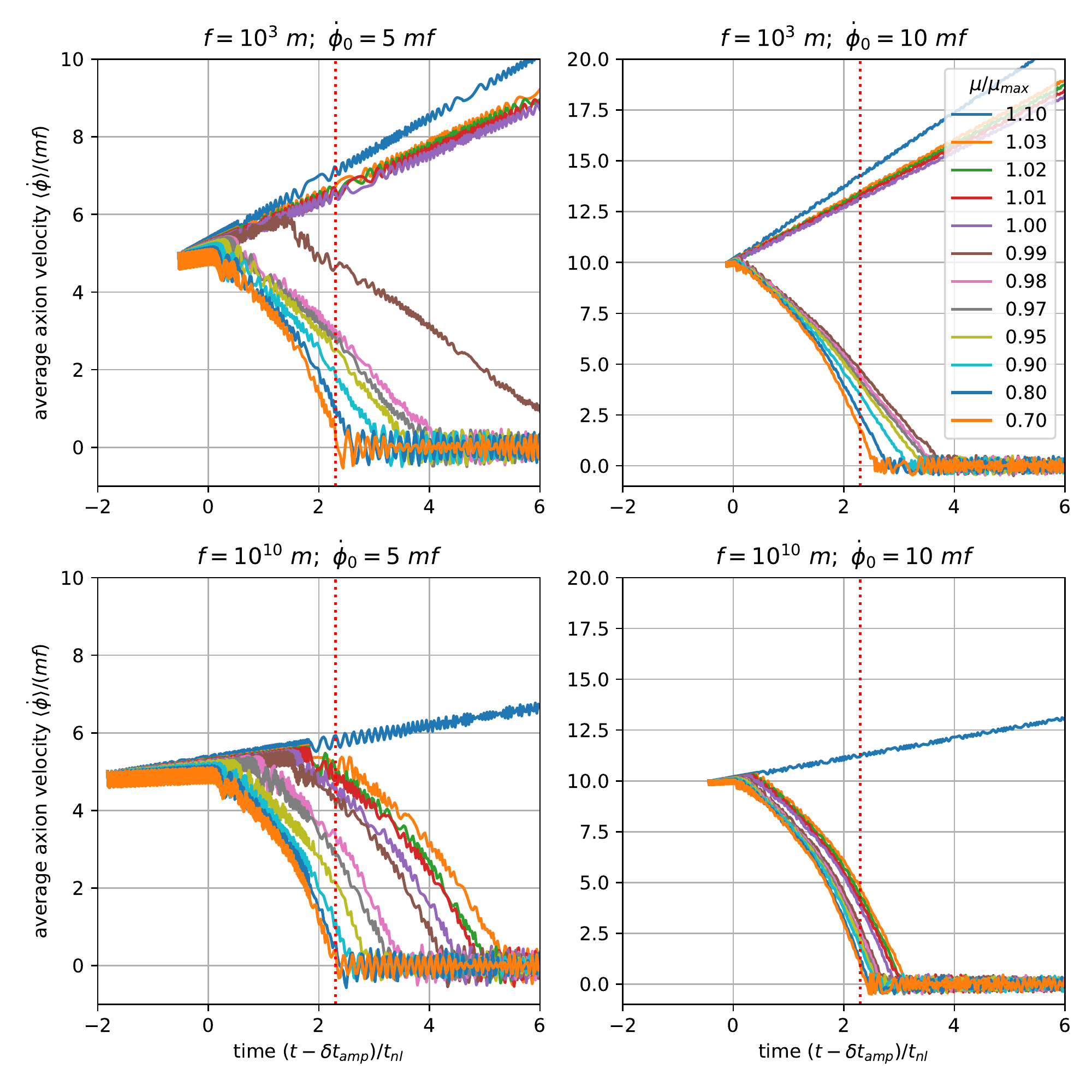}
\caption{\label{fig:mumax}
Average axion velocity varying the slope in the range $0.70 < \mu/\mu_{max} < 1.10$, for $f / m = 10^3$ (top) and $10^{10}$ (bottom), and $\dot\phi_0 = 5 m f$ (left) and $10 m f$ (right). The red dotted line is at $(t-\delta\tamp)/\tnl = 2.3$.
}
\end{figure}

The last quantity that we want to compute on the lattice is the maximal slope of the potential $\mu_\mathrm{max}$, which is defined from Eq.~(\ref{eq:condition for no positive phiddot}) with $H=0$:

\begin{equation}
\mu^3 < \mu_\mathrm{max}^3 \equiv \frac{\pi\Lambda_b^8}{2f\dot\phi_0^2} \left( W_0\left( \displaystyle\frac{32\pi^2f^4}{e\dot\phi_0^2} \right) \right)^{-1}\,.
\label{eq:mumax}
\end{equation}
For $\mu>\mu_\mathrm{max}$, fragmentation is not efficient enough to contrast the acceleration induced by the potential slope. Fig.~\ref{fig:mumax} shows the evolution of the zero mode for $\mu$ around $\mu_\mathrm{max}$, for different values of $f$ and of the initial velocity. It can be seen that the maximal value of $\mu$ for which the field stops respects Eq.~(\ref{eq:mumax}) with a percent accuracy.

\bigskip

In the closing of this section, let us briefly comment on the effect of the Hubble friction. Contrary to the slope term, the Hubble friction acts to slow down the rolling of $\phi$. When the Hubble friction is the dominant source of the friction, the fluctuation in $\phi$ remains small enough to use the linear analysis shown in \cite{Fonseca:2019ypl}. In this regime, the two sources of the friction can be written as
\begin{align}
\left( \frac{d\rho}{dt} \right)_{\rm frag} &= -\frac{\dot\phi^3|\ddot\phi + H\dot\phi|}{32\pi^2 f^4} \exp\left( \frac{\pi\Lambda_b^8}{2f\dot\phi^2|\ddot\phi + H\dot\phi|} \right), \label{eq:drhodt frag}\\
\left( \frac{d\rho}{dt} \right)_{\rm Hubble} &= - 3H\dot\phi^2. \label{eq:drhodt hubble}
\end{align}
For the derivation of Eq.~(\ref{eq:drhodt frag}), see Ref.~\cite{Fonseca:2019ypl}.
As long as $|(d\rho/dt)_{\rm frag}| \ll |(d\rho/dt)_{\rm Hubble}|$, the fragmentation effect is not important
and the time evolution of the zero mode is described by the equation of motion $\ddot\phi + 3 H \dot\phi - \mu^3 - (\Lambda_b^4/f) \sin(\phi/f) = 0$.
The fragmentation effect becomes important when $|(d\rho/dt)_{\rm frag}| \gtrsim |(d\rho/dt)_{\rm Hubble}|$, which occurs for
\begin{align}\label{eq:Hubble condition frag}
H \lesssim {\cal O}(1) \times \frac{\pi \Lambda_b^8}{f\dot\phi^3} \log \frac{32\pi^2 f^4}{\dot\phi^2}.
\end{align}
Here, we assumed $\mu^3 \lesssim {\cal O}(1) \times H\dot\phi$ otherwise Eq.~(\ref{eq:condition for no positive phiddot}) is not satisfied and $\phi$ keeps rolling.
Once this condition is satisfied, $(d\rho/dt)_{\rm frag}$ quickly dominates over $(d\rho/dt)_{\rm Hubble}$ because of the exponential factor.
Thus, we conclude that Hubble friction is not important once the fragmentation starts, but it controls \textit{when} this happens. For an ALP rolling down its potential, fragmentation starts after $H$ drops below the RHS of Eq.~(\ref{eq:Hubble condition frag}). In the case of the relaxion, fragmentation starts as soon as the barriers appear, if Eq.~(\ref{eq:Hubble condition frag}) is satisfied.
This justifies our choice of not including cosmic expansion in our lattice simulations.

\section{Secondary fluctuations}\label{sec:NLO}
Secondary fluctuations will be sourced as higher order terms in the potential become important once the initial fluctuations in the resonance band have grown. While our lattice analysis takes these effects into account to all orders, we here first outline the approach of calculating them to second order analytically and afterwards compare to the lattice.\\

To capture the secondary fluctuations, we extend the linear ansatz from Eq.~(\ref{eq:ansatz}) by a second order term
\begin{equation}
 \phi(x,t)=\phi(t)+\delta\phi(x,t)+\delta^{(2)}\phi(x,t).
\end{equation}
The second order fluctuations $\delta^{(2)}\phi$ are of $\mathcal{O}(\delta\phi^2)$ and initially zero. Plugging this ansatz into the full equation of motion, going to Fourier space, and separating the $\mathcal{O}(\delta\phi^0)$ and $\mathcal{O}(\delta\phi^1)$ pieces we find Eq.~(\ref{eq:fluctuation}) and also an equation for the $\mathcal{O}(\delta\phi^2)$ terms in the limit of vanishing expansion
\begin{equation}
 \ddot{\delta^{(2)}\phi_k}+(k^2+V''(\phi))\ \delta^{(2)}\phi_k=-\frac12 V'''(\phi) \int \frac{d^3p}{(2\pi)^3}\delta\phi_{p}\delta\phi_{k-p}=:S_k \,,\label{eq:NLO_ansatz}
\end{equation}
which is just the equation of a sourced harmonic oscillator. The particle physics interpretation of this result is, that higher order terms in the potential cause scattering of two axions in the excited modes $p$ and $k-p$ into an axion with momentum $k$.  Using the analytic results for the modes in the resonance band during the period were the backreaction is negligible and the axion therefore rolls with constant velocity Eq.~(\ref{eq:asymptotic uk}), the energy spectrum in the second order fluctuations becomes 
\begin{align}
 \frac{d\rho^{(2)}}{d\log k}
& \approx
\frac{k^2 \delta k_{cr}}{2^9\pi^3} \frac{\Lambda_b^8}{f^6}\frac{1}{t} \exp\bigg(4\delta k_{cr}t\bigg)  \theta(2k_{cr}-k)
\bigg[ \frac{1}{k^2+4\delta k_{cr}^2}+\frac{1}{(k-2k_{cr})^2+4\delta k_{cr}^2} \nonumber \\
& + \frac{1}{(k+2k_{cr})^2+4\delta k_{cr}^2} +\frac14 \frac{1}{(k-4k_{cr})^2+4\delta k_{cr}^2} + \frac14 \frac{1}{(k+4k_{cr})^2+4\delta k_{cr}^2} \bigg]\,,\label{eq:NLOspec}
\end{align}
as calculated in App.~\ref{sec:NLO_details}.
In the case of a narrow resonance defined by $\delta k_{cr} /\kcr \ll 1$,  the first and second term in the square brackets of Eq.~(\ref{eq:NLOspec}) correspond to secondary resonances at $k=0$ and $k=2\kcr$.
Notice that Eq.~(\ref{eq:NLOspec}) does not predict any resonance at $k = 4 k_{cr}$,  due to the finite $k$ range encoded in the $\theta$ function. The non-resonant terms are sizeable away from the resonance though, and we included them for completeness.

\begin{figure}
\centering
\includegraphics[width=\textwidth]{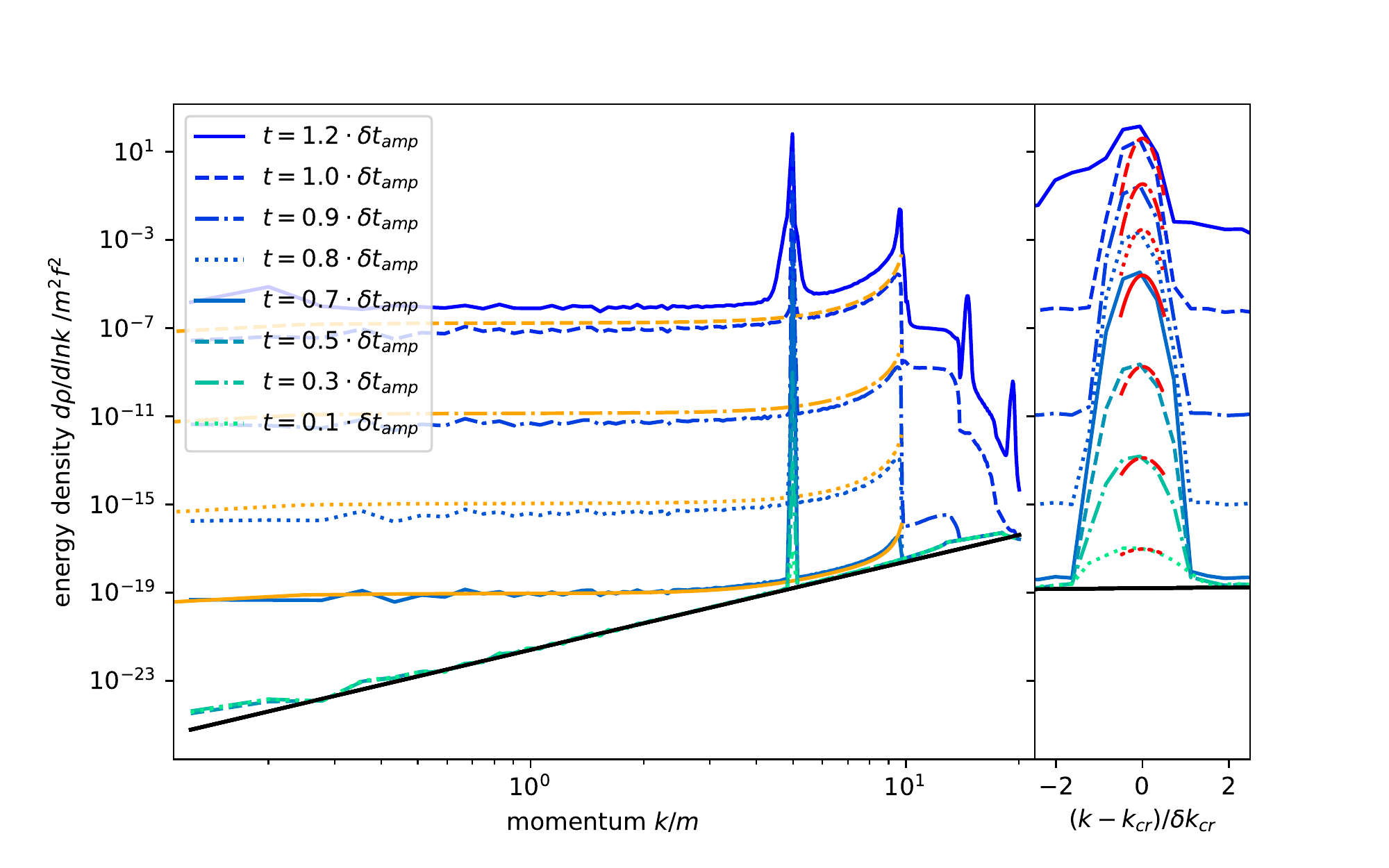}
\caption{\label{fig:EarlySpecs}
Early evolution of the axion energy spectrum for $f/m=10^{10},~\dot\phi_0=10 mf$. The blue shaded lines show the spectrum as obtained from a lattice with $N=512^3$ sites and side length $L=40/m$. The bottom black line is the analytic expression for the initial Bunch Davies vacuum $\propto k^4$ and the orange lines give the analytic NLO estimate Eq.~(\ref{eq:NLOspec}) for $t=0.7-1.0\cdot \tamp$. On the right we magnified the region around the peak $k_{\rm cr}=5 m$ and show for comparison the analytic LO estimate Eq.~(\ref{eq:asymptotic uk}) for $t=0.1-1.0\cdot \tamp$ in red. 
}
\end{figure}

The first two dominating contributions predict a flat spectrum at low momenta $2\delta k_{cr} \lesssim k \lesssim 2 \kcr$, and a secondary peak at $k = 2 \kcr$ corresponding to collinear scattering processes.
This expectation is indeed confirmed in Fig.~\ref{fig:EarlySpecs}, where we show the axion spectrum as obtained on the lattice for different times. Initially, the axion is taken to be in the Bunch-Davies vacuum shown in black at the bottom of the plot. On the right side of the plot we show a close up of the resonance band around $k_{cr}$. The exponential growth of the modes in the resonance band with time up to $t\approx\tamp$ is clearly visible as expected from the analytical result Eq.~(\ref{eq:asymptotic uk}) (shown in red for comparison). Around $t=0.7 \,\tamp$, the energy in the modes with $k<2k_{cr}$ starts growing at approximately twice the rate of the modes in the resonance band. These are the secondary fluctuations that arise as axions in the resonance band scatter in $2\rightarrow 1$ processes. The analytic estimate of this effect in Eq.~(\ref{eq:NLOspec}), shown in orange, predicts the order of magnitude as well as the main features of the spectrum accurately. As $t$ approaches $\tamp$, the energy in higher momentum modes is amplified as well.  The secondary peak at $k = 2 \kcr$ predicted by Eq.~(\ref{eq:NLOspec}) is clearly visible, as well as the primary one at $k = \kcr$. We believe that the additional peaks at higher momenta are due to higher order effects that eventually lead to the breakdown of perturbation theory. 

Perturbation theory fully breaks down at $\tamp$ when the axion zero mode slows down and the resonance band moves to smaller momenta. The new starting point for the amplification of the modes in the resonance band is not the initial spectrum anymore, but the sum of the initial spectrum and the secondary fluctuations. The time it takes for the energy in the modes to grow sufficiently to slow down the axion zero mode is therefore reduced and the axion stops faster, as we observed in Sec.~\ref{sec:lattice}. This also explains why the stopping process becomes independent of the initial spectrum after $\tamp$: if the initial perturbations are smaller than the induced secondary ones, they are simply negligible after this point.

\begin{figure}
\centering
\includegraphics[width=\textwidth]{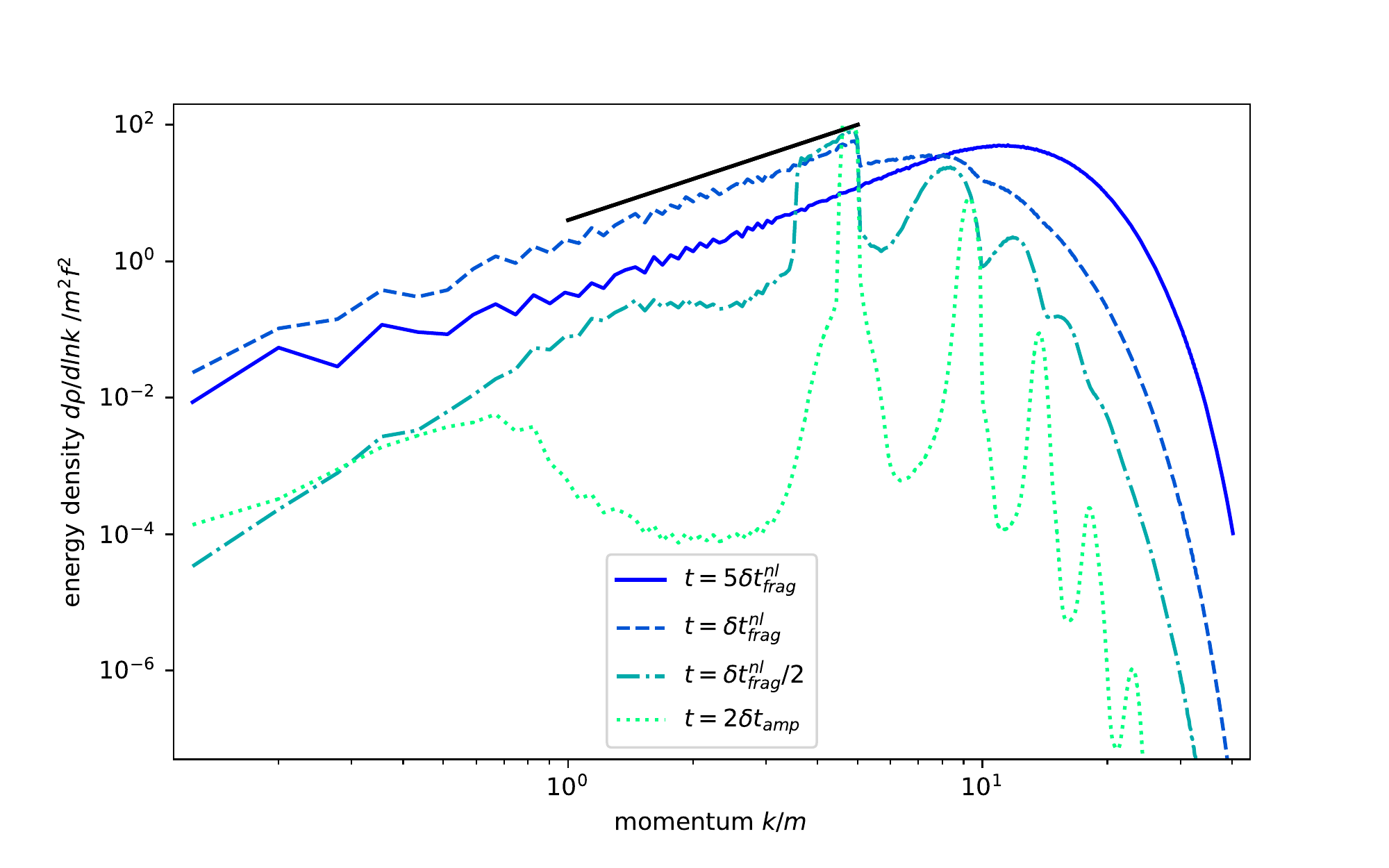}
\caption{\label{fig:LateSpecs}
Evolution of the axion energy spectrum past $\delta\tamp$ for $f/m=10^{10},~\dot\phi_0=10 mf$. The blue shaded lines show the spectrum as obtained from the same lattice as in Fig.~\ref{fig:EarlySpecs}.
The black line shows the spectrum from the linear analysis given in Eq.~(\ref{eq:final_spec}).
}
\end{figure}

In Fig.~\ref{fig:LateSpecs} we show the further evolution of the spectrum.
Again it is useful to come up with an expectation in the linear picture to be able to compare to the lattice and understand the effect of higher order processes. In the linear analysis, we can derive a simple analytic formula for the energy spectrum $d\rho/d\log k$. As the axion loses its kinetic energy, the resonance band sweeps from its initial position $\kcr=\dot\phi_0/2f$ to $\kcr=m$, when the axion gets trapped in the wiggles. Assuming the axion deposits its energy only into the resonance band, energy conservation tells us that $\int_{\dot\phi/2f}^{\dot\phi_0/2f} dk\ d\rho/dk = \dot\phi_0^2/2 - \dot\phi^2/2$. Then, we obtain \cite{Fonseca:2019ypl}

\begin{align}
 \frac{d\rho}{d\log k}=4f^2k^2\qquad \text{for}\qquad m<k<\frac{\dot\phi_0}{2f}.\label{eq:final_spec}
\end{align}
This estimation from the linear analysis is shown as the solid black line in Fig.~\ref{fig:LateSpecs}. We see that as the axion slows down, the spectrum is well matched by this estimate for modes with momenta bigger than the current critical momentum. We notice that the simulated spectrum is an $\mathcal{O}(1)$ factor smaller than the estimate. This can be easily understood, since higher order processes keep shuffling the energy into high momentum modes. The spectrum resulting from these processes is clearly visible for modes with $k>\dot\phi_0/(2f)$. Once the axion has stopped at $\Delta t^{\rm nl}_{\rm frag}$, there is no further energy injected into the axion inhomogeneities. The scattering processes however remain active and result in the peak of the spectrum moving to higher momenta.
Such an energy cascade into modes with higher momenta can be understood as the early state of the thermalization \cite{Boyanovsky:2003tc, Micha:2004bv, Destri:2004ck}.

\section{Formation of bubbles}\label{sec:bubbles}
A very important point that needs to be discussed is the possibility that the axion field populates multiple minima in spatially separated regions. 
If the fragmentation process takes place during inflation, these multiple minima would not be observable as the corresponding regions are stretched by the exponential expansion, and thus in the currently visible Universe the vacuum would be unique (unless fragmentation takes place during the last $\mathcal{O}(60)$ e-folds of inflation, in which case the discussion below applies).
On the other hand, if fragmentation takes place after inflation,  multiple minima can be populated within one Hubble patch. This scenario has multiple consequences, which we list here:
\begin{itemize}
\item First of all,  if multiple minima are populated, we expect a bubble wall structure to develop. Even if the dynamics is such that the field quickly relaxes to one single minimum within a Hubble volume, the selected minimum need not be the same in different Hubble patches. Hence, as the horizon grows and previously separated patches enter into causal contact, we expect at least one domain wall with an area $\sim H^{-2}$ to be present at any given time in the visible Universe. Depending on its energy, this may be problematic as it could lead to overclosure. This is indeed the case for the self-stopping relaxion, see Sec.~\ref{sec:relaxion bubbles}.
\item Secondarily, due to the overall slope of the potential $-\mu^3$,  different vacua have different vacuum energies.  If the energy difference is small, this could lead to an inhomogeneous cosmological constant (CC). If instead the energy difference is large, this would worsen the CC problem in that a fine tuning would be required for the different vacua to average at the correct value.
\item Finally, in the case of the relaxion, large spatial inhomogeneities of the field $\phi$ would lead to a inhomogeneous value of the Higgs VEV. We mention this here for completeness, but we do not expect it to be problematic as the differences in the electroweak VEV would be tiny by construction.
\end{itemize}
Even though the above possibilities are interesting by themselves, and may be viable depending on the parameters of the model, we will here assume that they do not occur, and compute the necessary conditions to avoid them. In particular, inhomogeneities may be created on three different length scales, which need to be analyzed separately.

\subsection{Fluctuations on super-Hubble scales}
If the axion is light compared to the Hubble scale during inflation, then it will be excited with a nearly scale invariant spectrum. Due to these fluctuations, we expect patches of the universe with different initial values of the axion field, meaning the axion velocity will also differ at the point when the wiggles in the axion potential appear and fragmentation stops the field shortly after. As we can see from Eq.~(\ref{eq:delta phi frag nl}), different initial velocities result in the fragmentation process stopping the field at different positions. If these differences are larger than the fundamental period $2\pi f$, this leads to the field stopping in different minima and therefore the existence of superhorizon bubbles. Even if dynamics eventually smooth the field value across the Hubble volume, as the horizon grows more regions in which the field has settled in different minima will enter into causal contact. Therefore, we expect to have multiple minima populated at any time within the visible Universe.

We expect inflation at a scale $H_I$ to result in approximately scale-invariant fluctuations with amplitude $\delta\phi\sim H_I/(2\pi)$ in the field before the scanning process begins.
If the height of the barriers does not depend on $\phi$ (as e.g. for generic ALPs),
$H_I \lesssim 2\pi f$ should be imposed to avoid domain wall formation.
On the other hand, if the height of the barriers does depend on $\phi$ (as in the Graham-Kaplan-Rajendran (GKR) relaxion model \cite{Graham:2015cka}), the constraint on $H_I$ is relaxed
because the fragmentation starts only when $\phi$ reaches the critical point where the Higgs VEV becomes non-zero and the barriers appear, leading to a reduction in the fluctuations in $\phi$.
In this case the following bound on the inflationary scale can be derived (see App.~\ref{sec:relaxion cosmo})
\begin{align}\label{eq:HImax}
 H_I\lesssim\frac{\pi^2}{z_{\phi}}\frac{\Lambda_b^8}{\dot\phi_0^4}\ \phiscan \,,
\end{align}
in order to avoid superhorizon bubbles in the case where the axion constitutes a subdominant component of the total energy. This is the case if the distance the axion rolls while scanning is sub-Planckian, \textit{i.e.,} $\phiscan\lesssim m_{pl}$. 
The bound in Eq.~(\ref{eq:HImax}) is mild,  especially when compared to the original GKR mechanism.  As shown in Fig.~\ref{fig:parameter space},  $H_I$ can be as large as $10^{16}\GeV$.  In the original GKR relaxion mechanism instead, it can never exceed $\mathcal{O}(10^2)\GeV$ and it is typically sub-GeV, or even as low as the meV range~\cite{Fonseca:2019lmc}.
In the case where the axion dominates the total energy and drives inflation or at least a period thereof, this bound disappears since fluctuations in the axion become equivalent to adiabatic fluctuations rather than isocurvature ones.

\subsection{Critical bubbles}
It is useful at this point to take a closer look at the different infrared scales involved in our setup. Regarding the bubbles, we follow Ref.~\cite{Lalak:2007rs} to estimate the width of the bubble wall at rest by minimizing the surface tension, \textit{i.e.}, the energy per unit wall area.  While the surface tension arising from the field being displaced from the minimum of the potential grows for larger bubble widths, the tension due to the gradient of the field is reduced. With these considerations, one finds the following estimates for the wall width $w$ and the surface tension $\sigma$
\begin{align}
 w\approx5 \, m^{-1}\,, \qquad \sigma\approx10\, mf^2 \,.
\end{align}
Notably, the scales where most of the energy is deposited are smaller than $m^{-1}$ and therefore smaller than the width of a bubble wall. The dynamics of these fluctuations therefore do not resemble the ones of bubbles and we discuss their impact in the next section. Furthermore, one can calculate the critical radius $R_{\text{crit}}$ a bubble needs to reach such that the pressure from the non-degeneracy of the vacua driving the expansion of the bubble overcomes the surface tension. 
\begin{align}
 R_{\text{crit}}\approx\frac{mf}{\mu^3} \,.
\end{align}
The question we would like to answer in this section is whether bubbles with radii bigger than $R_{\text{crit}}$ are formed in the stopping process. Those bubbles would keep expanding and it is uncertain whether such a system would finally settle in one common minimum. Unfortunately, it is impossible to answer this question with lattice simulations alone for the following reason: When we choose $\dot\phi_0=\mathcal{O}(10)\, mf$, such that the field is able to overcome the barriers initially, we need a lattice spacing $\Delta x\approx \mathcal{O}(10^{-2}-10^{-1}) \, m^{-1}$ in order to resolve the UV dynamics properly. Since current computing power only allow for simulations with $\mathcal{O}(10^{3})$ lattice sites along each spatial direction, it is impossible to also include $R_{\text{crit}}$, which in general is much larger than $m^{-1}$ even when choosing $\mu\approx\mu_{\text{max}}$ in Eq.~(\ref{eq:mumax}). We therefore highlight below two observations that we can make on the lattice and extrapolate to argue why there are no expanding bubbles.

\begin{figure}\centering
 \includegraphics[width=\textwidth]{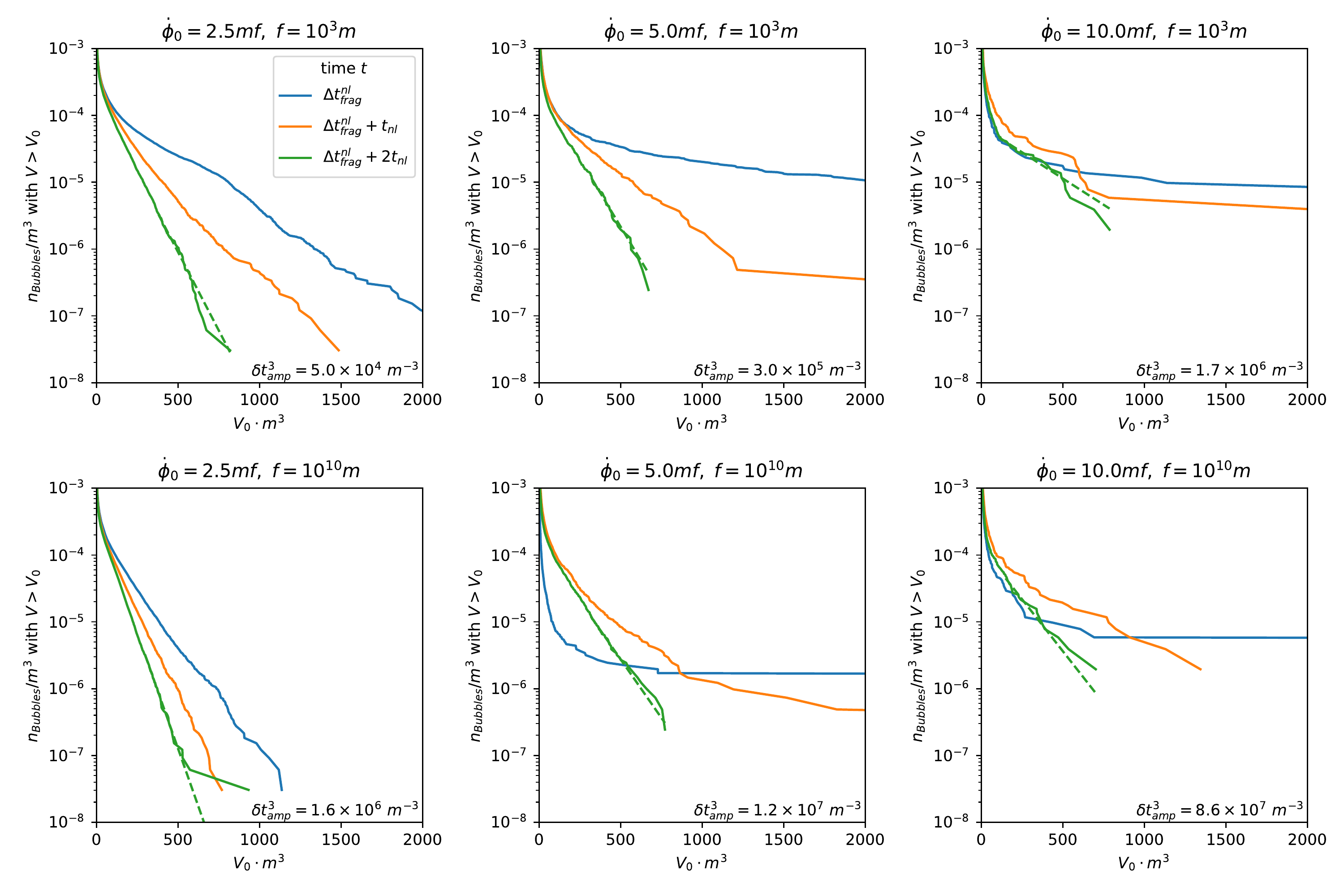}
 \caption{Dependence of the number density of bubbles with a volume bigger than $V_0$ for different $f$ and $\dot\phi_0$ at three different times. The dashed lines show the fit of an exponential decay $n(V_0)\propto\exp(-\Gamma V_0)$ to the last few data points for each time.}
 \label{fig:ExpBubbleSuppression}
\end{figure}

Our first observation is that when counting the number of bubbles exceeding a certain volume $V_0$ once the field has stopped rolling, we find that the number density of such bubbles is exponentially suppressed as one raises $V_0$. This is shown in Fig.~\ref{fig:ExpBubbleSuppression}. It becomes clear, however, that the details of this suppression are very complicated since they show a dependence on time as well as the parameters $\dot\phi_0$ and $f$. Additionally, especially for simulations with large initial velocities $\dot\phi_0$, the simulated box cannot be too large without compromising the resolution of the UV physics of fragmentation, resulting in poor statistics for very large bubbles. This being said, we note that the critical volume $R_{\text{crit}}^3$ is much larger than the volumes testable on the lattice and in the case of the relaxion where $\dot\phi_0\gg mf$, we also have $R_{\text{crit}}^3\gg m^{-3}$ such that we can expect the probability of an expanding bubble forming in the visible universe to be suppressed by a huge exponential factor.

\smallskip

The second argument,  which holds for bubbles of slightly larger size,  is based on the fact that in parts of space that are separated by more than the time of the first exponential amplification $\tamp$ or even the full time it takes the axion to stop $t_{\rm frag}$, the stopping processes are (partially) independent. They can be viewed as different instances of the same experiment, in which the observable is the rate of energy transfer to the field fluctuations or, equivalently,  the minimum in which the field ends up.

\begin{figure}\centering
 \includegraphics[width=0.8\textwidth]{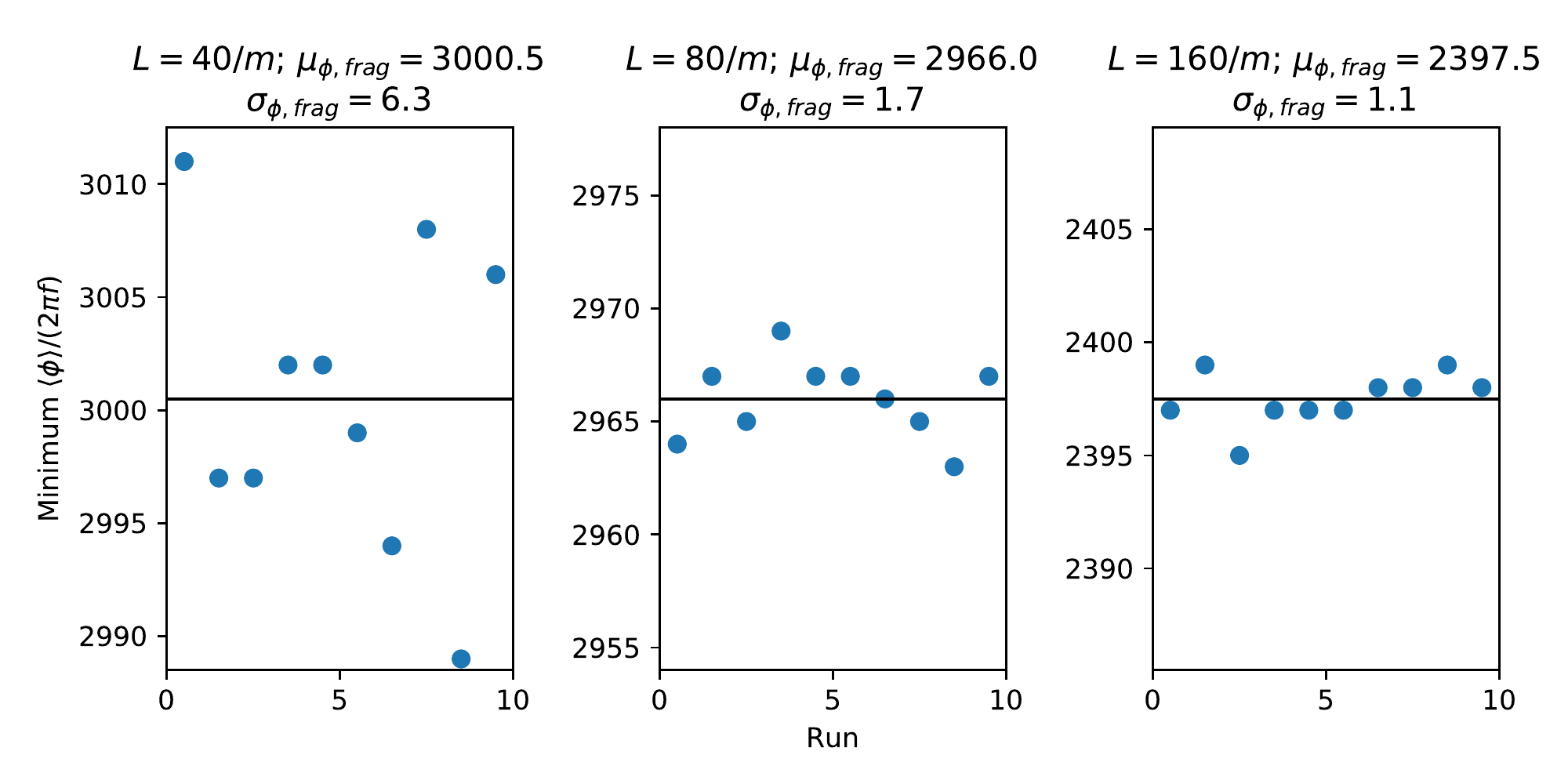}
 \caption{Spread of the minima the field stops in for $f=10^{10}m$ and $\dot\phi_0=10mf$ in different lattice configurations. The $y$-axis is centered around the average final minimum $\mu_{\phi,\text{frag}}$. Importantly, the spread of the minima $\sigma_{\phi,\text{frag}}$ decreases as the size of the box $L$ is increased.}
 \label{fig:MinSpread}
\end{figure}
If the field ends up in different minima in parts of space where the process takes place independently, we are left with bubbles at the end, as in the super-Hubble case. Such a situation can be avoided if the field average after fragmentation is the same at each instance of the quantum experiment. To check if this is the case, we ran 10 simulations with the same physical parameters $m,f,\dot\phi_0$ in boxes with increasing volume $V=L^3$. As Fig.~\ref{fig:MinSpread} shows, the spread of minima the field stops in $\sigma_{\phi,\text{frag}}$ reduces as the size of the box is increased. To check whether large, possibly expanding bubbles might exist after the field has stopped we need to extrapolate this result to infrared scales.  To do so, we estimate the variance of the total field excursion $\phifrag$. We assume that this is entirely due to the variance of $\tamp$ and the corresponding field excursion $\dot\phi_0\tamp$.
In App.~\ref{sec:variance tamp}, we analytically estimate the standard deviation in a box of size $\tamp$ (which is the smallest of the infrared scales $\tamp,\ R_{\text{crit}}$ and $t_{\rm frag}$) to be
\begin{equation}\label{eq:stdev tamp corrected}
\frac{\sigma_{\phi,\text{frag}}}{2\pi f}
\approx \mathcal{O}(10) \times \left[\log\left(\frac{8\pi f^2}{\dot\phi_0}\right)\right]^{-3/2} \,,
\end{equation}
where the multiplicative factor $ \mathcal{O}(10)$ is added to match the normalization of the analytical formula with the lattice calculation.
For the self-stopping relaxion, this quantity ranges roughly between $0.1$ and $0.01$.  This means that for a volume $(c \, \tamp)^3$,  different minima occur only at the $10-100 \sigma$ level.  This number cannot be simply translated into a probability, because we do not know the probability distribution to such an accuracy. If it were Gaussian, the probability would be between $10^{-22}$ and $10^{-2200}$.  One of course would have to impose that this very rare occurrence does not happen in any of the small volumes that constitute our Universe. Not knowing the actual probability distribution, performing such a calculation is not illuminating, thus we content ourselves with imposing $\sigma_{\phi,\text{frag}} / (2\pi f) \ll 1$ in Eq.~(\ref{eq:stdev tamp corrected}).

\subsection{Small scale fluctuations}
While the majority of the energy is dumped into fluctuations on scales $\leq m^{-1}$, these fluctuations are on scales too small to be thought of as bubbles, since they are smaller than the typical width of a bubble wall of $\mathcal{O}(m^{-1})$. They do however interfere with the previously discussed fluctuations on larger scales, in that they cause a spread of the field. If this spread is comparable or larger than one period of the axion potential $2\pi f$, the dynamics on large scales and of the mean field become less sensitive to the potential. 
We will argue below, however, that the spread in the axion field is always smaller than $2\pi f$ (although not by much) such that the expected corrections have only a minor influence on the discussion above.

We can estimate the spread of the field by using the analytic final energy spectrum in Eq.~(\ref{eq:final_spec}), which for relativistic modes $k>m$ results in the following power spectrum
\begin{align}
 P_{\phi}(k)=\frac{1}{k^2}\frac{d\rho}{d\log k}=4f^2\qquad \text{for}\qquad m<k<\frac{\dot\phi_0}{2f}.
\end{align}
Integrating this spectrum, we find that the root-mean-square (RMS) of the axion field is given by
\begin{align}
\delta \phi_{\rm rms} = \sqrt{\langle\delta\phi^2\rangle}=2f\left(\log{\frac{\dot\phi_0}{2mf}} \right)^{1/2} \,,
\label{eq:axion_RMS}
\end{align}
so the spread of the field is indeed comparable to the period of the potential, but very high initial velocities would be required for it to be bigger due to the square root and logarithmic dependence. In the specific case of the relaxion, we find the square root to be in the range $0.2 - 2$,  implying that the fluctuations on this scale are indeed smaller than $2\pi f$.

\begin{figure}
 \includegraphics[width=0.95\textwidth]{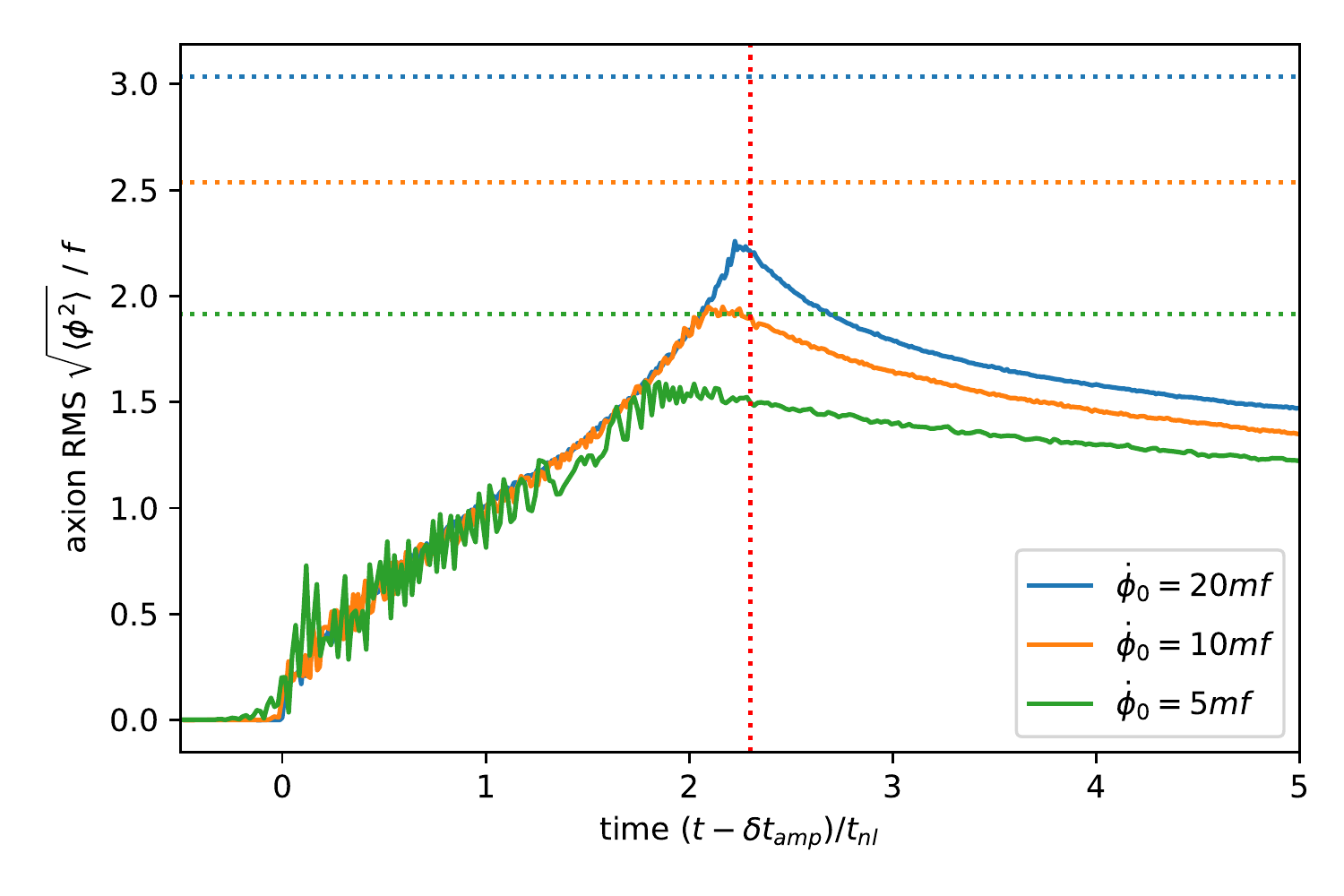}
 \caption{Evolution of the axion RMS field value caused by modes with $k>m$ as computed by integration of the axion power spectrum obtained from a lattice computation with $N=256^3$ lattice sites, $L=40/m$, and $f=10^{10}m$ (solid lines). The dotted horizontal lines show the analytic estimate from Eq.~(\ref{eq:axion_RMS}). The horizontal, dotted, red line marks the time around which the axion stops as estimated in Sec.~\ref{sec:lattice}.  }
 \label{fig:axion_RMS_evolution}
\end{figure}

In Fig.~\ref{fig:axion_RMS_evolution}, we show the evolution of $\delta\phi_{\rm rms}$ as the field stops as computed by integration over the modes with $k>m$ in the axion power spectrum obtained from the lattice. We see that $\delta\phi_{\rm rms}$ starts growing significantly around the time when the production of axion fluctuations starts to slow down the axion zero mode (around $0$ with the chosen normalization of the $x$-axis) and reaches its maximum around the time when the axion stops and no more energy is transferred into axion fluctuations (red vertical line). We note that the maximal $\delta\phi_{\rm rms}$ is smaller than the analytic estimate in Eq.~(\ref{eq:axion_RMS}) (given by the dotted horizontal lines), and that it further decreases after the axion has stopped rolling. Both of these effects can be attributed to the higher order effects discussed in Sec.~\ref{sec:NLO}, since the scattering of axions redistributes the energy into higher momentum modes in the non-linear regime. In the relativistic case, the energy density and the power spectrum are related by a factor $k^2$, so this leads to a reduction in the integral over the power spectrum (which gives the mean square of the field), while the integral over the energy spectrum is conserved as it must be. In an expanding universe, one additionally has a depletion of the energy, so this effect would be pronounced even more.

Note that the amount by which the analytic result overestimates the peak of the numerical result grows with the initial velocity $\dot\phi_0$, signaling that the actual dependence of $\delta\phi_{\rm rms}$ on the initial velocity is even weaker than predicted by the analytic estimate. We therefore conclude that the spread of the axion field is smaller than $2\pi f$ for a wide range of initial velocities, such that our previous considerations are not significantly affected by the small scale fluctuations.

\section{Relaxion considerations}
\label{sec:relax_consids}

In this section, we discuss implications from our lattice results to the relaxion mechanism.

\subsection{Relaxion bubbles}\label{sec:relaxion bubbles}
One of the most interesting features of fragmentation as a stopping mechanism is that the relaxion mechanism does not need strong Hubble friction and therefore the relaxation phase does not need to take place during inflation~\cite{Fonseca:2019lmc}.  The main advantage of a post-inflationary relaxation phase is that some of the issues which are typically associated with the embedding of the relaxion into inflation disappear. In particular, the number of e-folds does not need to be exceedingly large, but can as well be $\mathcal{O}(10-100)$, and the Hubble rate during inflation can be much larger, being only constrained by Eq.~(\ref{eq:HImax}).

However, when relaxation takes place after inflation, there is the possibility of forming relaxion bubbles, \textit{i.e.}, spatially separated patches in which the relaxion field ends up in different minima as discussed in Sec.~\ref{sec:bubbles}. The existence of such bubbles would have the following consequences: First, since the Higgs VEV depends on the relaxion field, the EW scale would have slightly different values in each of these regions. This does not seem problematic since a variation in $\phi$ of size $\Delta \phi \sim 2\pi f$ corresponds to a tiny difference in $\vEW$ by construction of the relaxion mechanism.
There is however an apparent problem tied to the fact that the difference in potential energy from one minimum of the relaxion potential to the next, namely $2 \pi f g \Lambda^3$,  is much larger than the measured value of the cosmological constant $\mathcal{O}(10^{-47})\GeV^4$. Therefore, even if one assumes that the average value of the CC matches the observed one, the CC would be unacceptably inhomogeneous. We therefore assume that such bubbles do not form, under the criteria derived in Sec.~\ref{sec:bubbles}.
In addition,  the scenario discussed in Sec.~\ref{sec:bubbles} would result in the presence of at least one domain wall of area $H^{-2}$ at any time.  The energy density of such an object (given the relaxion parameters) would overclose the universe, which is another reason to impose Eq. ~(\ref{eq:HImax}).

\subsection{Higgs fluctuations}\label{sec:Higgs}
The full potential in the case where $\phi$ is identified as the relaxion field necessarily includes couplings to the Higgs in order to scan the Higgs mass as well as trigger barriers when the Higgs acquires a non-zero VEV. The required potential can be written as
\begin{equation}
V(\phi, h) = \Lambda^{4} - g\Lambda^{3} \phi + \frac{1}{2}(\Lambda^{2} - g' \Lambda \phi)h^2 + \frac{\lambda}{4}h^4 + \Lambda_{b}^{4} \frac{h^2}{v_{\rm EW}^{2}} \cos \frac{\phi}{f} \,.
\end{equation}
As the relaxion rolls over many fundamental periods, the effective Higgs mass
\begin{equation}
\frac{\partial^2 V}{\partial h^2} = \Lambda^{2} - g' \Lambda \phi + 3 \lambda \langle h^2 \rangle + 2 \left(\frac{\Lambda_b^2}{v_{\rm EW}}\right)^2  \cos \frac{\phi}{f} \,,
\end{equation}
is a rapidly oscillating function, leading to an instability that amplifies fluctuations of the Higgs field. Following the analysis of Sec.~\ref{sec:linearsummary}, there is an instability band for
\begin{equation} 
\frac{\dot{\phi}^2}{4f^2} - \frac{\Lambda_b^4}{v_{\rm EW}^2} < k^2 + m_{\rm eff}^2 < \frac{\dot{\phi}^2}{4f^2} + \frac{\Lambda_b^4}{v_{\rm EW}^2} \,,
\end{equation}
with $m_{\rm eff}^2= \Lambda^{2} - g' \Lambda \phi + 3 \lambda \langle h^2 \rangle \equiv m_{h}^2(\phi) +3 \lambda \langle h^2 \rangle $. Initially, the Higgs mass $m_{h}^2(\phi) \sim \Lambda^2$ is large and positive so there is no instability and we have $\langle h^{2} \rangle = 0$. However, as the relaxion field scans the potential, the effective Higgs mass decreases and modes will begin to enter the resonance band and grow exponentially. In turn, the quartic induced, effective mass $\propto \langle h^{2} \rangle$ grows until the mode again exits the instability band. This interplay between the decrease in effective mass due to the evolution of the relaxion field and the increase due to the quartic induced mass leads to a so-called edge solution where the mode stays fixed at the upper edge of the instability band~\cite{Ibe:2019udh}. Once the edge solution is established, the zero mode obeys the condition
\begin{equation}
m_{h}^2(\phi)+ 3 \lambda \langle h^2 \rangle =  \frac{\dot{\phi}^2}{4f^2} + \frac{\Lambda_b^4}{v_{\rm EW}^2}  \,,
\end{equation}
meaning that the typical energy in the Higgs field is
\begin{align}
\rho_h \sim \lambda \langle h^2 \rangle^2 = \frac{1}{9\lambda} \left( \frac{\dot\phi^2}{4f^2} + \frac{\Lambda_b^4}{v_{\rm EW}^2} - m_h^2(\phi)  \right)^2.
\end{align}

In order to see the effect of the Higgs fluctuations, let us estimate the energy of the Higgs field during the last stage of relaxation where we have $0\lesssim m_h^2(\phi) \lesssim v_{\rm EW}^2$.
First, we note that $\Lambda_b \lesssim \sqrt{4\pi} v_{\rm EW}$ is typically expected in simple UV completions, see \textit{e.g.,} App.~A of Ref.~\cite{Fonseca:2019lmc}.
Therefore, if $\dot\phi / f \ll v_{\rm EW}$ is satisfied when the edge solution is established, then $\rho_h$ is at most of the order $v_{\rm EW}^4$, meaning that the Higgs field cannot absorb a large fraction of the total relaxion kinetic energy and the oscillation of the Higgs zero mode is negligible compared to the Higgs VEV.
This condition is indeed satisfied in the most of the viable self-stopping relaxion parameter space previously identified in Ref.~\cite{Fonseca:2019lmc}, meaning that the effect of Higgs fluctuations is small compared to the friction from relaxion fragmentation.
Moreover, the regulated growth of the Higgs field due to the quartic leads to an edge solution which is strictly less efficient than the unregulated exponential growth of relaxion fluctuations during the scanning phase. We thus conclude that while growth of Higgs field can occur, it does not significantly alter the success of the self-stopping relaxion mechanism, nor its parameter space. 

On the other hand, if $\dot\phi / f \gtrsim v_{\rm EW}$, the amplitude of Higgs zero mode can be larger than $v_{\rm EW}$ before relaxation completes. In this case, the analysis of the relaxation process should involve both the relaxion and the Higgs field, and the stopping condition should be modified. This scenario is interesting, but beyond the scope of this paper. Here, we will simply assume that the condition $\dot\phi / f \ll v_{\rm EW}$ is satisfied and show its impact on the viable self-stopping relaxion parameter space in Sec.~\ref{sec:relaxion params}.

\subsection{Parameter space}\label{sec:relaxion params}

In this section, we want to briefly discuss how the parameter space of the relaxion is modified once the new conditions discussed in this paper are taken into account.  For a thorough discussion of all the conditions that the model has to satisfy, we refer the reader to Ref.~\cite{Fonseca:2019lmc}.
There are two modifications with respect to this discussion. First, the lattice simulation of Sec.~\ref{sec:lattice} and the second order calculation of Sec.~\ref{sec:NLO}  show that fragmentation is more efficient than the purely linear expectation.  Second, in order to avoid the growth of Higgs fluctuations, we have to add the condition $\dot\phi / f \ll v_\ew$ as discussed in Sec.~\ref{sec:Higgs}.

Concerning the first point, we proceed as in Ref.~\cite{Fonseca:2019lmc}. There, the parameter space was derived by using Eqs.~(\ref{eq:mastertimescale}) and (\ref{eq:masterfieldexcursion}),  and replacing $\log(\ldots) \to 50$. Analogously, the product log in Eq.~(\ref{eq:condition for no positive phiddot}) was replaced by $W_0(\ldots) \to 50$.
Here we proceed analogously by keeping Eqs.~(\ref{eq:mastertimescale}),~(\ref{eq:masterfieldexcursion}),~(\ref{eq:condition for no positive phiddot}) but now we replace $\log(\ldots) \to 2$ to account for the shorter stopping time found in the lattice analysis. However, we keep $W_0(\ldots) \to 50$ as in~\cite{Fonseca:2019lmc}, because Eq.~(\ref{eq:condition for no positive phiddot}) concerns the onset of fragmentation, which occurs when the fluctuations are still in the linear regime and hence the linear analysis is still valid.

\begin{figure}
\centering
During inflation (Sec. 3.1 of \cite{Fonseca:2019lmc})\\
\includegraphics[height=.4\textwidth]{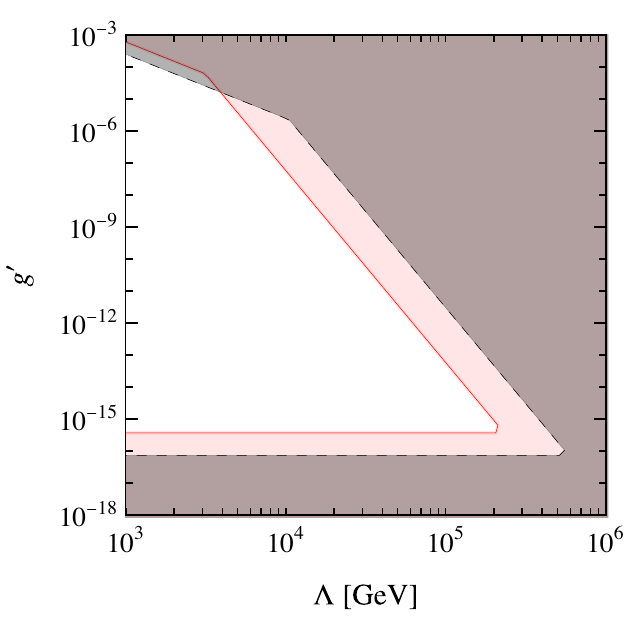}
\includegraphics[height=.4\textwidth]{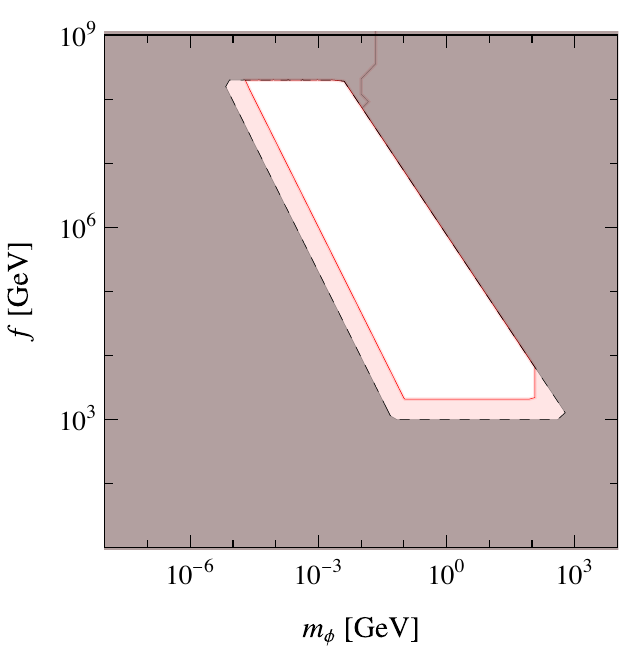}\\
\vspace{3mm}
After inflation,  $g/g' = 1$ (Sec. 3.2 of \cite{Fonseca:2019lmc})\\
\includegraphics[height=.4\textwidth]{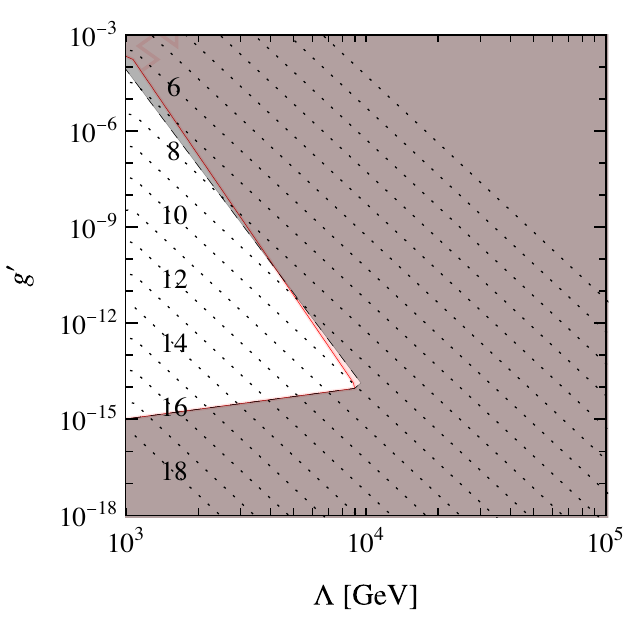}
\includegraphics[height=.4\textwidth]{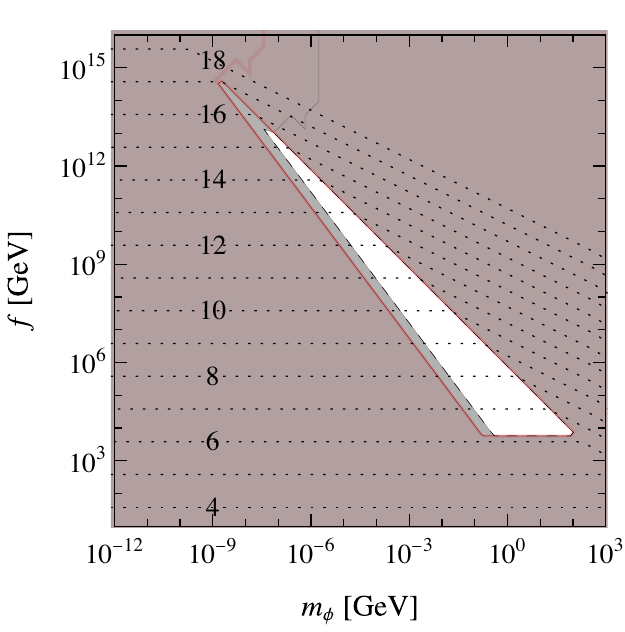}\\
\vspace{3mm}
After inflation,  $g/g' = 1/(4\pi)^2$ (Sec. 3.2 of \cite{Fonseca:2019lmc})\\
\includegraphics[height=.4\textwidth]{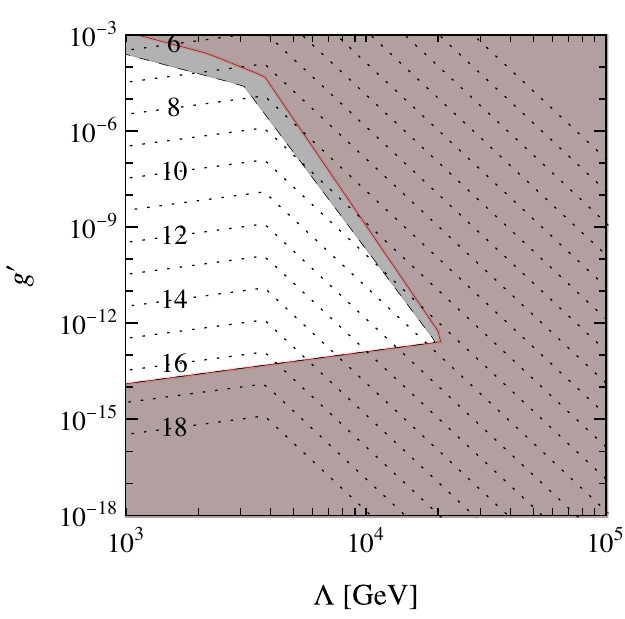}
\includegraphics[height=.4\textwidth]{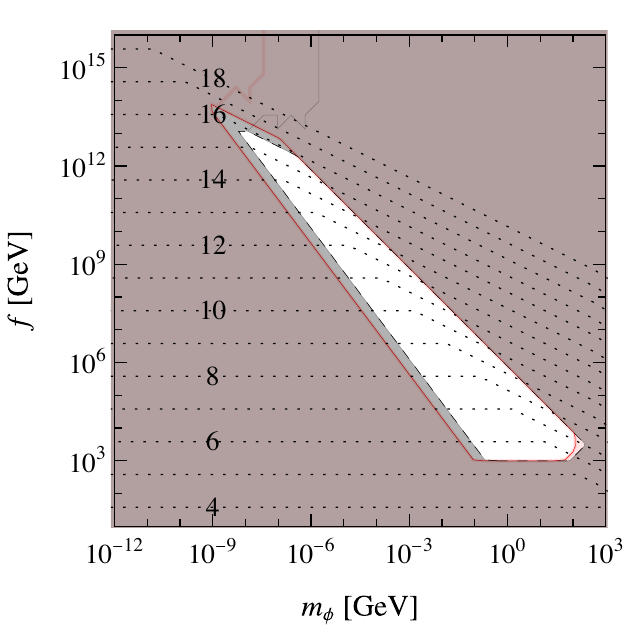}
\caption{Parameter space of the relaxion model including the results of this paper (in red), compared to the results of~Ref.~\cite{Fonseca:2019lmc} (in gray, dashed lines).  Top: Relaxation during inflation. Center: Relaxation after inflation, with $g/g' = 1$. Bottom: Relaxation after inflation, with $g/g' = 1/(4\pi)^2$.  In the center and bottom rows, we superimpose the contours of $\log_{10} H_I^\mathrm{max}$, defined according to Eq.~(\ref{eq:HImax}).
}
\label{fig:parameter space}
\end{figure}

In Fig.~\ref{fig:parameter space}, we show a comparison of the parameter space of Ref.~\cite{Fonseca:2019lmc} (in gray, dashed lines) with that of this work (in red),  for three reference scenarios. In the top row, we consider the case of relaxation during inflation. In the center and bottom rows, relaxation takes place after inflation. For this latter case, we superimpose the contours of the maximal allowed value of $H_I$,  according to Eq.~(\ref{eq:HImax}).
We fix $g/g'$ as in Ref.~\cite{Fonseca:2019lmc}, while all other parameters are left free to vary. 
We see that the new results of this paper lead to a slight reduction in the viable parameter space of the self-stopping relaxion model.

\section{Conclusion}
In this work, we have analyzed axion fragmentation using a classical lattice simulation.
We have confirmed that the kinetic energy of the axion zero mode dissipates into fluctuations in a manner similar to the expectations of  Ref.~\cite{Fonseca:2019ypl}, with some important modifications coming from non-linearities that can only be captured by the lattice simulation. As shown in Fig.~\ref{fig:lattice params}, one such modification is that the dissipation of the zero mode kinetic energy is even more efficient compared to the linear approximation used in Ref.~\cite{Fonseca:2019ypl} because modes outside of the instability band are populated due to $2\to 1$ processes. These secondary fluctuations dominate over the initial fluctuations and thus enhance the dissipation effect in the non-perturbative regime. This is an NLO effect and therefore not included in the analysis of Ref.~\cite{Fonseca:2019ypl}, but is captured to all orders in our lattice simulation. Moreover, since the amplitude of the secondary fluctuations is determined by $2\to 1$ processes, the fragmentation process in non-perturbative regime is insensitive to the particular choice of the initial spectrum of fluctuations as shown in Fig.~\ref{fig:flat vs Bunch Davies}.

In Sec.~\ref{sec:bubbles}, we have discussed the fluctuations after the completion of fragmentation as well as bubble/domain wall formation.
Since the typical size of the fluctuation $\langle \delta\phi^2\rangle$ is of the order of $f^2$, one might worry about problematic domain wall formation. We therefore determined the conditions such that the dynamics of axion fragmentation do not result in domain walls of cosmological size, and we showed that they mainly concern the initial condition of the axion evolution,  which are set during inflation, and hence the inflationary Hubble scale $H_I$.

Finally, in Sec.~\ref{sec:relax_consids}, we examined the consequences of bubble formation as well as the possible excitation of Higgs fluctuations in the relaxion specific case. Bubble formation in the case of the relaxion leads to unacceptable cosmology and thus must be avoided by imposing an upper bound on the scale of inflation. Regarding Higgs fluctuations, we argue that in a large fraction of the viable parameter space for the self-stopping relaxion, the Higgs field cannot absorb a large fraction of the relaxion's kinetic energy and thus the growth of Higgs fluctuations has only a minor impact on the viable parameter space. The new constraints on the parameter space, including the enhanced dissipation of the relaxion's kinetic energy due to NLO effects, are discussed in Sec.~\ref{sec:relaxion params}.

In summary, we have shown directly via lattice simulation that fragmentation is a very efficient mechanism of depleting kinetic energy from an axion field rolling over many oscillations of a periodic potential. In the special case where the axion is identified as the self-stopping relaxion, we have quantified the parameter space where fragmentation as a stopping mechanism leads to successful relaxation of the electroweak scale.

\section*{Acknowledgements}

We thank Cem Er\"oncel, Keisuke Harigaya, Ken'ichi Saikawa, G\'eraldine Servant, Philip S\o rensen, and Motoo Suzuki for useful discussions.
The work of EM and WR is supported by  the  Cluster  of  Excellence  ``Precision  Physics,  Fundamental  Interactions,  and  Structure of Matter'' (PRISMA+ EXC 2118/1) funded by the German Research Foundation (DFG) within the German Excellence Strategy (Project ID 39083149).

\clearpage

%------------------------------
\appendix
%------------------------------
\section{NLO Calculation}
\label{sec:NLO_details}
Here we give the details for how to obtain the NLO spectrum Eq.~(\ref{eq:NLOspec}) from the Ansatz Eq.~(\ref{eq:NLO_ansatz}):
\begin{equation}
 \ddot{\delta^{(2)}\phi_k}+(k^2+V''(\phi))\ \delta^{(2)}\phi_k=-\frac12 V'''(\phi) \int \frac{d^3p}{(2\pi)^3}\delta\phi_{p}\delta\phi_{k-p}=:S_k \,.
\end{equation}
We start by noting that the $V''$ term on the left averages to zero and is therefore only relevant for the modes in the resonance band as long as the axion is rolling. These modes are dominated by the first order perturbations anyhow and we therefore drop the $V''$ from now on. The equation can then be formally solved to give
\begin{equation}\label{eq:2nd order soln}
 \delta^{(2)}\phi_k(t)=\int^t_{t_i} dt'\ \frac{\sin(k(t-t'))}{k}S_k(t') \,,
\end{equation}
with $t_i\to -\infty$.  The energy density in these modes is then given as
\begin{align}\label{eq:rho}
\langle\rho(x,t)\rangle & =\left \langle \frac{1}{2} (\dot{\delta^{(2)}\phi} )^2 + (\vec\nabla \delta^{(2)}\phi)^2 \right\rangle \\
& = \frac12 \int \frac{d^3k}{(2\pi)^3}  \frac{d^3k'}{(2\pi)^3} e^{-i (\vec k+\vec k') \vec x} \langle \delta^{(2)}\dot\phi_{\vec{k}} \delta^{(2)}\dot\phi_{\vec{k'}} + \vec k \cdot \vec k' \, \delta^{(2)}\phi_{\vec{k}} \delta^{(2)}\phi_{\vec{k'}} \rangle
\end{align}
where $\langle\dots\rangle = \langle 0 | \dots | 0 \rangle$.  By plugging Eq.~(\ref{eq:2nd order soln}) into (\ref{eq:rho}) one obtains
\begin{equation}\label{eq:drhodlogk}
\frac{d\rho}{d\log k} = \frac{k^3}{4\pi^2} \int^t_{t_i} dt'dt'' \cos(k (t'-t'')) S^2(k, t', t'') \,,
\end{equation}
where we defined the unequal time correlator (UTC) $S^2(k,t',t'')$ as
\begin{align}
 \langle0|S_{\vec k}(t') S^*_{-\vec k'}(t'')|0\rangle=(2\pi)^3\delta^{(3)}(\vec k + \vec k')S^2(k,t',t'') \,.
\end{align}
When the axion rolls with a constant velocity $\dot\phi_0=2fk_{cr}$ the source reads
\begin{equation}
 S_k(t)=-\frac{\Lambda_b^4}{f^3}\sin(2k_{cr} t)\int \frac{d^3p}{(2\pi)^3}\delta\phi_p \delta \phi_{k-p}.
\end{equation}
In the following we are going to consider the case in which the fluctuations in the resonance band are initially in Bunch-Davies vacuum $\delta\phi_k(t)=a_k u_k(t)+a_{-k} u_{-k}^\dagger$ with the mode functions $u_k(t)$ given by Eq.~(\ref{eq:asymptotic uk}) for concreteness. 
When calculating the vacuum expectation value it turns out that only the following combination contributes for finite momenta $k=k'\neq0$
\begin{align}
 \bra0 a_{{p}} a_{{k}-{p}} a^\dagger_{{p'}}a^\dagger_{{k'}-{p'}}\ket0=(2\pi)^6[\delta^{(3)}({k}-{p}-{p'})+\delta^{(3)}({p}-{p'})]\delta^{(3)}({k}-{k'})
\end{align}
and we find for the UTC
\begin{align}
 S^2(k,t',t'')=\frac{\Lambda_b^8}{f^6}\sin(2k_{cr}t')\sin(2k_{cr}t'')\int \frac{d^3p}{(2\pi)^3}\ 2\ u_p(t')u_{k-p}(t')u_{p}^*(t'')u_{k-p}^*(t'').
\end{align}
Since the mode functions only depend on the absolute momentum, we evaluate the momentum integral choosing $|\vec p|$ and $|\vec k-\vec p|$ as our integration variables, together with a trivial angular integration, since the problem is invariant under rotations around $\vec k$
\begin{align}
 \int \frac{d^3p}{(2\pi)^3}
 &=\frac{1}{(2\pi)^2}\int_0^{\infty}dp\int_{|k-p|}^{k+p} dq\ \frac{pq}{k}.
\end{align}
The mode functions are sharply peaked around $k=k_{cr}$ and can be approximated as Gaussian in the peak region
\begin{align}
 u_k(t)\approx \frac{1}{\sqrt{2k_{cr}}}\exp\bigg(\delta k_{cr}t-\frac{(k-k_{cr})^2}{2\delta k_{cr}}t\bigg)\sin\bigg(k_{cr}t+\frac{\pi}{4}\bigg).
\end{align}
For $k>\sqrt{\delta k_{cr}/(t'+t'')}$ the Gaussian peak lies fully within the momentum integration then and we find
\begin{align}
 \begin{split}
 S^2(k,t',t'')&=\frac{1}{4\pi}\frac{\Lambda_b^8}{f^6} \frac{\delta k_{cr}}{k(t'+t'')} \cdot\\
  &\quad\bigg[\exp\bigg(2\delta k_{cr}t'\bigg)\bigg(\frac14+ \frac12 \sin(2k_{cr}t')-\frac14 \cos(4k_{cr}t')\bigg)\bigg]\cdot \bigg[t'\rightarrow t''\bigg].
 \end{split}
\end{align}
When we plug this expression back into the equation for the energy density (\ref{eq:drhodlogk}), all we are left with are the two time integrals.  Due to the time-dependent exponential,  the integral is dominated by the region $t',t''\approx t$. We therefore replace $t'+t''$ in the numerator above by $2t$ and expand $\cos(k(t'-t''))$, which allows us to factorize the two integrals.  The integration can be then done explicitly.  After dropping all oscillating terms, which have frequencies $2n k_{cr}$,  with $n = 1, \dots, 4$, we arrive at Eq.~(\ref{eq:NLOspec}). 

\section{Relaxion Cosmology and super-horizon Bubbles}
\label{sec:relaxion cosmo}
In this section we discuss the evolution of perturbations in the relaxion field prior to the fragmentation. We start with the case in which the relaxation takes place after inflation and only comprises a subdominant fraction of the total energy density. For simplicity we assume that the universe is filled with a fluid with constant equation of state parameter $w>-1$. In this case one can choose the time coordinate such that the Hubble is given as
\begin{align}
 H(t)=\left(\frac{3}{2}(1+w)t\right)^{-1}.
\end{align}
The relaxion's zero-mode equation of motion in an expanding universe is 
\begin{align}
 \ddot\phi+3H\dot\phi+V'(\phi)&=0.
\end{align}
While the relaxion scans the Higgs mass we have $V'=-\mu^3$. We are going to assume that around $\phi=0$ the correct Higgs mass is reached, wiggles in the potential appear and the relaxion stops shortly after. The solution to the relaxion's equation of motion is then given as
\begin{align}
 \phi(t)&=-\phiscan+\frac{\mu^3}{2}\frac{1+w}{3+w}t^2=-\phiscan+\frac{\mu^3}{H^2(t)}\frac{2}{9(1+w)(3+w)}\\
 \dot\phi(t)&=\mu^3\frac{1+w}{3+w}t=\frac{\mu^3}{H(t)}\frac{2}{3(3+w)}
\end{align}
where $\phiscan$ is the distance the relaxion has to traverse in order to scan the Higgs mass. The Hubble when fragmentation takes place and the initial velocity are then given by 
\begin{align}
 H_0&=\sqrt{\frac{2}{9(1+w)(3+w)}\frac{\mu^3}{\phiscan}}\\
 \dot\phi_0&=\sqrt{\frac{2(1+w)}{(3+w)}\mu^3\phiscan}.
\end{align}
One can easily check that for $\phiscan\lesssim m_{pl}$ the relaxion's contribution to the total energy density is indeed subdominant.

To see the effect of isocurvature fluctuations, let us now take the separate universe approach and consider a patch, where the distance the field has to roll is modified by a fluctuation $\phiscan\rightarrow \phiscan+\delta\phi$. In this patch the scanning process takes longer because the field has to traverse a bigger distance, which will lead to a smaller Hubble when $\phi=0$ as well as a bigger velocity.

\begin{align}
 \delta H_0&=-H_0 \frac{\delta \phi}{2\phiscan}\\
 \delta\dot\phi_0&=\dot\phi_0\frac{\delta\phi}{2\phiscan}\label{eq:dotphipert}
\end{align}

Once fragmentation starts, Hubble friction is negligible and the relaxion stops in a fraction of a Hubble time. The effect of the perturbation to the Hubble while fragmentation is active is therefore negligible. The difference in the initial velocity, however, leads to the field rolling further $\Delta \phi_\mathrm{frag}\rightarrow\Delta \phi_\mathrm{frag}+\delta \phi_\mathrm{frag}$, as can be estimated using Eq.~(\ref{eq:delta phi frag nl}). 
\begin{equation}
 \delta \phi_\mathrm{frag}\simeq 4\Delta \phi_\mathrm{frag}\frac{\delta\dot\phi_0}{\dot\phi_0}=2\Delta \phi_\mathrm{frag}  \frac{\delta\phi}{\phiscan}\,,
\end{equation}
where we assumed that the field excursion during the initial amplification is negligible as is the case for the parameter space discussed in~\cite{Fonseca:2019lmc}. Using that the fluctuations on super-Horizon scales caused by inflation are given by $\delta\phi=H_{I}/(2\pi)$ and that the fluctuations after stopping should not exceed $\pi f$ in order to avoid super-Horizon bubbles, we arrive at Eq.~\ref{eq:HImax}.

\section{Variance of $\phifrag$}
\label{sec:variance tamp}

According to Eq.~(\ref{eq:delta t frag nl}), the time required for fragmentation to complete $\tfrag$ can be split into a an interval $\tamp$, in which the quantum fluctuations of the axion, with momenta inside the initial instability band, get exponentially enhanced and classicalized, and another interval $\tnl \cdot z_t$ in which the evolution is dominated by higher order scattering processes. The first interval lasts until the instability band, whose position depends on the zero-mode velocity, moves to the IR by an amount equal to its initial width. This can be determined by using energy conservation, and depends on the initial energy of the modes within the instability band. The latter quantity, which we denote by $E_0$, is a quantum observable, the variance of which will determine the variance of $\tamp$.
We find it reasonable to assume that the variance of $\tfrag$, and correspondingly $\phifrag$, can be entirely ascribed to the variance of $\tamp$, since after this point the process proceeds classically and its duration is fixed by the spectrum within the instability band at $\tamp$.

The time $\tamp$ is determined as follows.  The energy in the instability band increases as
\begin{equation}
\delta E = E_0 \exp (2 \deltakcr \tamp ) - E_0 \approx  E_0 \exp (2 \deltakcr \tamp )
\end{equation}
In this time interval, the instability band moves by $-2\deltakcr$, thus the variation of the kinetic energy $K$ is
\begin{equation}
\delta K = - \frac{d K}{d k_\mathrm{cr}} 2\deltakcr = 2 \dot\phi_0^2 \frac{\deltakcr}{k_\mathrm{cr}} = 2 \Lambda_b^4 \,.
\end{equation}
Energy conservation implies
\begin{equation}
\tamp = \frac{1}{2\deltakcr} \log \left(\frac{2\Lambda_b^4}{E_0} \right) \,.
\end{equation}
Within this interval, the field evolves by an amount $\tamp \dot\phi_0$.
Computing the variance, in the probabilistic sense, of $\tamp$ is complicate task. Here, we will limit ourselves to compute its variation assuming $E_0$ changes by one standard deviation $\sigma_{E_0}$:
\begin{equation}\label{eq:variation tamp}
\Delta (t_{amp}) \approx \left| \frac{d  \tamp}{d E_0} \right| \sigma_{E_0} =  \frac{1}{2\deltakcr} \frac{\sigma_{E_0}}{E_0}.
\end{equation}
Now we need to compute $E_0$ and $\sigma_{E_0}$.  $E_0$ is the expectation value of the initial energy density, obtained recalling that in the Bunch-Davies vacuum $\operatorname{E}[|u_k|^2] = 1/(2k)$
\begin{align}
E_0 & = \int \frac{d^3 k}{(2\pi)^3} k^2 \operatorname{E}[|u_k|^2]  = \frac{4\pi k_\mathrm{cr}^4 (2\deltakcr)}{(2\pi)^3} \operatorname{E}[|u_k|^2]  = \frac{k_\mathrm{cr}^3 \deltakcr}{2\pi^2}  = \frac{1}{32 \pi^2} \frac{\dot\phi_0^2 \Lambda_b^4}{f^4}
\end{align}
To compute $\sigma_{E_0}$, we need to know the variance of $u_k$.  $u_k$ is gaussianly distributed, $\mathcal{P}(u_k) \propto \exp(-2 k |u_k|^2)$. The modulus follows a Rayleigh distribution,  $\mathcal{P}(|u_k|) =4 k |u_k| \exp(-2 k |u_k|^2)$, thus
\begin{align}
\operatorname{E}[|u_k|^2] & = 1/(2k) \\
\operatorname{Var}[|u_k|^2] & = 1/(2k)^2
\end{align}
The process we are considering takes place in a finite time $\tamp$.  In this time, points is space separated by more than $c\cdot\tamp$ can not interfere with each other, hence we can think of enclosing our problem in a box of size $L = c\cdot\tamp$.  Momenta are thus discrete and given by
\begin{equation}
\vec k = \frac{2\pi}{L}\vec i \,,
\end{equation}
with $\vec i  = (i_1, i_2, i_3)$, and $i_k \in \mathbb{Z}$.
The number of modes inside the instability band $k_\mathrm{cr} - \deltakcr < k < k_\mathrm{cr} + \deltakcr$ is
\begin{equation}
N \approx \frac{4\pi k_\mathrm{cr}^2 (2\deltakcr)}{(2\pi / L)^3} \,.
\end{equation}
Now we can compute the variance, assuming that all modes have the same momentum and the same variance, which is valid for $2\pi/L \ll \deltakcr \ll k_\mathrm{cr}$:
\begin{align}
\operatorname{Var}[E_0] & = \operatorname{Var} \left\{\frac{1}{(2\pi)^3} \left(\frac{2\pi}{L}\right)^3 \sum k^2 |u_k|^2 \right\} \nonumber \\
& = \left[ \frac{1}{L^3} k_\mathrm{cr}^2 \right]^2 \operatorname{Var} \left[ \sum |u_{k_cr}|^2 \right] \nonumber \\
& = \left[ \frac{1}{L^3} k_\mathrm{cr}^2 \right]^2 N \operatorname{Var} \left[ |u_{k_cr}||^2 \right] \nonumber \\
& = \frac{1}{4 \pi^2 L^3} k_\mathrm{cr}^4 \deltakcr \,. \label{eq:varE0}
\end{align}
The standard deviation $\sigma_{E_0}$ is simply $(\operatorname{Var}[E_0])^{1/2}$.
Combining Eq.~(\ref{eq:varE0}) with (\ref{eq:variation tamp}),  we obtain
\begin{align}\label{eq:stdev tamp L}
\frac{\Delta(\dot\phi_0 \sigma_{\tamp})}{2\pi f} 
& = \frac{\dot\phi_0}{4 f L^{3/2} k_\mathrm{cr} \deltakcr^{3/2}} \,.
\end{align}
Finally,  we can plug in $L = \tamp$:
\begin{align}\label{eq:stdev tamp}
\frac{\Delta (\dot\phi_0 \tamp)}{2\pi f} & = \frac{\dot\phi_0}{4 f  k_\mathrm{cr} \deltakcr^{3/2}} (2 \deltakcr)^{3/2} \log\left(\frac{2\Lambda_b^4}{E_0}\right)^{-3/2} \nonumber \\
& = \frac12 \log\left(\frac{8\pi f^2}{\dot\phi_0}\right)^{-3/2}
\end{align}
In the parameter space of the self-stopping relaxion, this quantity ranges between $0.01$ and $0.001$ for $\dot\phi_0 = \Lambda^2 = (10^5)^2\,\mathrm{GeV}^2$ and $f$ up to $10^{10} \,\mathrm{GeV}$.

\begin{figure}
\centering
\includegraphics[width=.7\textwidth]{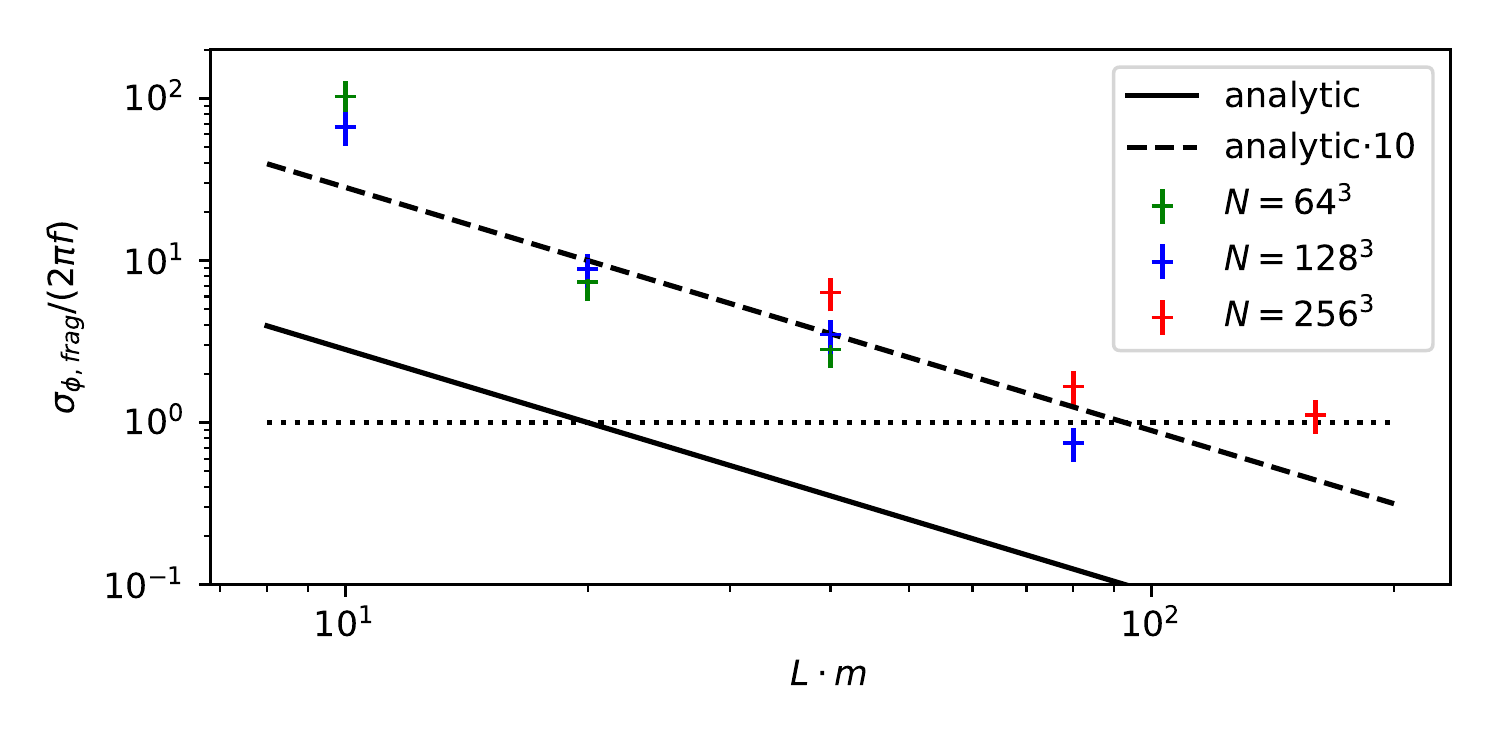}
\caption{\label{fig:phifrag variance comparison}
Dependence of the spread in the final position of the zero-mode of the field $\sigma_{\phi,\text{frag}}$ on the length of the sides of the simulated box $L$. The different color crosses represent simulations with different numbers of lattice sites,  all with $f=10^{10} m$ and $\dot\phi_0=10mf$. The solid line corresponds to the analytic estimate of Eq.~(\ref{eq:stdev tamp L}), which seems to underestimate the spread by $\mathcal{O}(10)$ but captures the decrease of the spread with increasing length $L$ correctly.}
\end{figure}

Checking Eq.~(\ref{eq:stdev tamp}) on the lattice is not easy, because the lattice size is typically smaller than $c\cdot \tamp$.  We can instead compare Eq.~(\ref{eq:stdev tamp L}) for a smaller box, of size $L$, with an estimate of the same quantity obtained by running multiple lattice simulations and computing the standard deviation of $\phifrag$. The result of such a comparison is shown in Fig.~\ref{fig:phifrag variance comparison}.  We can see that, for relatively small box sizes,  the estimate of Eq.~(\ref{eq:stdev tamp L}) underestimate the result by a factor of roughly $10$,  while the dependence on $L$ is compatible with the one obtained from the lattice.

\bibliographystyle{JHEP}
\bibliography{lattice_frag.bib}

\end{document}